\documentclass[letterpaper,titlepage,11pt]{article}

\pdfoutput=1

\usepackage{amssymb,amsmath,amsfonts}
\usepackage{epsfig}
\usepackage{ccaption}
\usepackage{graphicx}

\usepackage[
      colorlinks=true,
      linkcolor=blue,
      urlcolor=blue,
      filecolor=black,
      citecolor=red,
      pdfstartview=FitV,
      pdftitle={},
        pdfauthor={Roberto Emparan, Ryotaku Suzuki, Kentaro Tanabe},
        pdfsubject={},
        pdfkeywords={},
        pdfpagemode=None,
        bookmarksopen=true
      ]{hyperref}

\def\clock{{\count0=\time
           \divide\count0 60
           \ifnum\count0<10 0\fi\the\count0
           \multiply\count0 -60 \advance\count0 \time
           :\ifnum\count0<10 0\fi \the\count0
         }}
\newcommand{\timestamp}{{\small\vbox{\hbox{\tt\jobname.tex}
\hbox{\the\day/\the\month/\the\year, \clock}}}}

\newcommand{\ie}{{\it i.e.,\,}}
\newcommand{\eg}{{\it e.g.,\,}}

\newcommand{\lp}{\left(}
\newcommand{\rp}{\right)}

\newcommand{\mc}[1]{\mathcal{#1}}

\newcommand{\beq}{\begin{equation}}
\newcommand{\eeq}{\end{equation}}
\newcommand{\bea}{\begin{eqnarray}}
\newcommand{\eea}{\end{eqnarray}}
\newcommand{\beqa}{\begin{eqnarray}}
\newcommand{\eeqa}{\end{eqnarray}}

\newcommand{\R}{\mathbb{R}}

\newcommand{\N}{\mathbb{N}}
\newcommand{\sR}{\mathsf{R}}

\newcommand{\fr}[1]{\frac{1}{#1}}

\newcommand{\ord}[1]{{\mathcal O}(#1)}

\newcommand{\cS}{{\mathcal S}}

\newcommand{\cL}{{\mathcal L}}

\newcommand{\homega}{\hat{\omega}}
\newcommand{\hlam}{\hat{\lambda}}
\newcommand{\hatl}{\hat{l}}
\newcommand{\cD}{{\mathcal D}}

\newcommand{\nonum}{\nonumber\\ }

\numberwithin{equation}{section}
\begin{document}

\begin{titlepage}
%\timestamp
\rightline{YITP-14-45}
\rightline{KEK-TH-1741}
%\leftline{}{\timestamp}
\vskip 1cm
\centerline{\LARGE \bf Decoupling and non-decoupling dynamics}
\medskip
\centerline{\LARGE \bf of large $D$ black holes}
%\centerline{\LARGE \bf Black hole quasinormal modes at large $D$:}\medskip
%\centerline{\LARGE \bf Decoupling and non-decoupling spectra}
%\vskip 0.15cm
%\centerline{\Large \bf }
\vskip 1.2cm
\centerline{\bf Roberto Emparan$^{a,b,c}$, Ryotaku Suzuki$^{d}$, Kentaro Tanabe$^{b,e}$}
\vskip 0.5cm
\centerline{\sl $^{a}$Instituci\'o Catalana de Recerca i Estudis
Avan\c cats (ICREA)}
\centerline{\sl Passeig Llu\'{\i}s Companys 23, E-08010 Barcelona, Spain}
\smallskip
\centerline{\sl $^{b}$Departament de F{\'\i}sica Fonamental, Institut de
Ci\`encies del Cosmos,}
\centerline{\sl  Universitat de
Barcelona, Mart\'{\i} i Franqu\`es 1, E-08028 Barcelona, Spain}
\smallskip
\centerline{\sl $^{c}$Yukawa Institute for Theoretical Physics,}
\centerline{\sl  Kyoto University, Kyoto 606-8502, Japan}
\smallskip
\centerline{\sl $^{d}$Department of Physics, Osaka City University, Osaka 558-8585, Japan}
\smallskip
\centerline{\sl $^{e}$Theory Center, Institute of Particles and Nuclear Studies, KEK,}
\centerline{\sl  Tsukuba, Ibaraki, 305-0801, Japan}
\smallskip
\vskip 0.5cm
\centerline{\small\tt emparan@ub.edu,\, ryotaku@sci.osaka-cu.ac.jp,\, ktanabe@post.kek.jp}

\vskip 1.2cm
\centerline{\bf Abstract} \vskip 0.2cm 
\noindent 
The limit of large number of dimensions localizes the gravitational field of a black hole in a well-defined region near the horizon. The perturbative dynamics of the black hole can then be characterized in terms of states in the near-horizon geometry. We investigate this by computing the spectrum of quasinormal modes of the Schwarzschild black hole in the $1/D$ expansion, which we find splits into two classes. Most modes are \textit{non-decoupled modes}:  non-normalizable states of the near-horizon geometry that straddle between the near-horizon zone and the asymptotic zone. They have frequency of order $D/r_0$ (with $r_0$ the horizon radius), and are also present in a large class of other black holes. There also exist a much smaller number of
\textit{decoupled modes}: normalizable states of the near-horizon geometry that are strongly suppressed in the asymptotic region. They have frequency of order $1/r_0$, and are specific of each black hole. Our results for their frequencies are in excellent agreement with numerical calculations, in some cases even in $D=4$.

\end{titlepage}
\pagestyle{empty}
\small
%\tableofcontents
\normalsize
\newpage
\pagestyle{plain}
\setcounter{page}{1}

\section{Introduction}

In recent work we have advocated the use of a small parameter in the study of black hole physics, namely, $1/D$ when the number of spacetime dimensions $D$ is large \cite{Asnin:2007rw,Emparan:2013moa}. One important property of this limit is that black holes possess well-defined near-horizon regions with universal features and enhanced symmetry \cite{Emparan:2013xia}, affording analytical control over several problems in perturbative black hole dynamics \cite{Asnin:2007rw,Emparan:2013moa,Emparan:2013oza,Emparan:2014cia,Emparan:2014jca}. 

The existence of sharply defined near-horizon geometries is a familiar feature of charged or rotating black holes that are close to extremality. 
Such near-horizon regions, however, are not present for generic black holes away from any extremal limit, in particular for Schwarzschild black holes. These possess only one scale, the horizon radius $r_0$, and therefore all their dynamics --- \eg\ their free, unforced oscillations --- occurs over distances of the same order as $r_0$. However, when $D$ is regarded as a parameter that is allowed to be large, the strong localization of the gravitational field results in the appearance of a small length scale $r_0/D$ which determines the extent of a near-horizon region where all the non-trivial black hole physics takes place. 

In this article we study the implications of this phenomenon for the classical perturbative dynamics of the Schwarzschild black hole, specifically the quasinormal spectrum of its oscillations, which we compute in analytic form in the $1/D$ expansion. We investigate its characterization in terms of the dynamics of the near-horizon geometry, and find a sharp distinction between two classes of quasinormal modes:

\begin{enumerate}
\item Non-decoupling modes, with frequencies $\omega\sim D/r_0$, straddle between the near-horizon zone and the asymptotic region. In the near-horizon geometry they are non-normalizable states. Most quasinormal modes fall in this class. This spectrum carries little information about the black hole geometry and is in fact universally shared by asymptotically flat, static black holes.\footnote{For rotating black holes, this spectrum appears in boosted form by the rotation of the horizon \cite{Emparan:2014jca}.}

\item Decoupled modes, with $\omega\sim 1/r_0$,\footnote{Throughout this article $\omega=\mc{O}(1/r_0)$ means $\omega=\mc{O}(D^0/r_0)$, \ie\ in this regime we may have $\omega\gg 1/r_0$ as long as $\omega r_0$ is parametrically smaller than $D$.} and angular momentum number $\ell\ll D$, have wavefunctions strongly suppressed in the asymptotic region, and can be said to decouple from it. They are localized within the near-horizon zone, where they are normalizable states. These are few modes, and are specific of each black hole.

\end{enumerate}
This appearance of two different scalings with $D$ of the quasinormal frequencies has been first identified numerically in \cite{Dias:2014eua}.

The non-decoupling spectrum, with frequencies of the order of the surface gravity, is expected \cite{Emparan:2013moa}. But the existence of a decoupled sector of the dynamics at much lower frequencies is a surprise. In contrast to the long `throats' that appear in (near-)extremal black holes, the near-horizon region of the Schwarzschild black hole at large $D$ has very small radial extent, so one would not expect to find states trapped for an arbitrarily long time within it and decoupled from the asymptotic region.

Normally, the existence of decoupled dynamics requires that states of finite frequency, as measured in the near-horizon time scale, are  normalizable states within the near-horizon geometry. The precision about the time scale is important, since due to the small radial size of the large $D$ near-horizon region, the characteristic `near-horizon time' $\hat t$ runs $D$ times faster than the time $t=\hat t/D$ of the asymptotic region, and therefore a finite far-zone frequency $\omega$ is a vanishingly small frequency $\hat\omega=\omega/D$ when measured in near-horizon time scales.
 
Most of the quasinormal modes that we find have finite non-zero frequency $\hat\omega$ and are \textit{not} normalizable near the horizon, hence not decoupled from the asymptotic region.\footnote{Quanta of Hawking radiation also do not decouple: their typical frequencies are very high, $\sim D^2/r_0$ \cite{Hod:2011zzb}, and therefore leave easily the near-horizon region.} These modes are present for the three types of gravitational perturbations obtained according to their $SO(D-1)$ tensorial character, namely scalar-, vector- and tensor-type. Thus, there is a degeneracy $\propto D^2$ for every quasinormal mode of partial wave number $\ell$ and overtone number $k$. One interesting property of modes with $k\ll D$ is that their damping ratio vanishes,
\beq
\frac{\text{Im}\,\omega}{\text{Re}\,\omega}\sim D^{-2/3}\to 0\,,
\eeq
so these modes can be said to approach normal, non-dissipative oscillations \cite{Emparan:2014cia}. Higher overtone modes have damping ratios of order one or larger.

Interestingly, we also find a decoupled sector of black hole dynamics with very different properties. It consists of a few quasinormal modes of gravitational vector and scalar types, with $\ell=\mc{O}(1)$ and finite frequencies $\omega$ as measured in the slower asymptotic time $t$. Thus their near-horizon frequencies $\hat\omega=\omega/D$ vanish when $D\to\infty$. So, to leading order in $1/D$, these normalizable states are \textit{static} modes in the near-horizon geometry, which become dynamical only at the next order in $1/D$. Still, they remain decoupled at all perturbative orders in the expansion. These modes are not universal but instead depend on the specific black hole geometry beyond the leading large $D$ limit. Therefore they can encode features, such as stability properties \cite{Emparan:2014jca}, that distinguish among different neutral black holes with the same leading near-horizon geometry. Their damping ratios are of order one. The existence at leading large $D$ order of these static, zero-mode perturbations of the horizon dovetails with the observation in \cite{Emparan:2013moa} that when $D\to\infty$ black holes appear to become `soft', \ie\ arbitrarily deformable.

In addition to uncovering these aspects of the black hole spectrum at large $D$, our study also demonstrates the large $D$ expansion as a calculational tool. Some of our results can be checked for accuracy against the recent numerical computations of \cite{Dias:2014eua}. For non-decoupling modes with low overtone number $k$, we find that our analytical result for $\text{Re}\,\omega$ provides a good approximation to the numerical values even at moderate values of $D$. However, our calculation of $\text{Im}\,\omega$ for these modes appears to be accurate only at very high values of $D$. In this respect, the interest of the latter result is more formal than practical.
In contrast, in the decoupled sector, where we have obtained the frequencies up to terms  of order $1/D^3$ in the expansion, we find remarkably good agreement with the numerical calculations.

The plan of the paper is the following: in the next section we discuss the main qualitative aspects of the large $D$ limit of the effective radial potentials for the black hole perturbations, and of the near-horizon geometry. In sec.~\ref{sec:farnear} we solve the perturbation equations in the far- and near-zones, and find their respective forms in the overlap zone. In sec.~\ref{sec:nondec} we match them to obtain the quasinormal frequencies of non-decoupling modes, and in sec.~\ref{sec:decspec} those of the decoupled sector. In sec.~\ref{sec:numbers} we compare our results to the numerical calculations of \cite{Dias:2014eua}. Sec.~\ref{sec:omegaplane} gives a brief graphical summary of our findings, and we conclude in sec.~\ref{sec:fini} with some additional comments. In one of the appendices we resolve an issue posed in \cite{Emparan:2014cia} and show that the universal non-decoupling spectrum is also present in extremal charged black holes.

\section{Qualitative analysis of large $D$ black hole dynamics}
\label{sec:qualia}

\subsection{Effective radial potentials}

The main qualitative aspects of the quasinormal spectrum of Schwarzschild black holes at large $D$ can be anticipated from the form of the effective radial potential for the perturbations \cite{Kodama:2003jz}. In 
\beq
D=n+3
\eeq
spacetime dimensions, we consider the black hole solution \cite{Tangherlini:1963bw}
\beq\label{schtang}
ds^2=-f(r)dt^2+\frac{dr^2}{f(r)}+r^2 d\Omega_{n+1}
\eeq
with
\beq
f(r)=1-\frac{r_0^n}{r^n}\,,
\eeq
and study its linearized gravitational perturbations $\delta g_{\mu\nu}=e^{-i\omega t}h_{\mu\nu}(r,\Omega)$. The angular dependence can be separated and the perturbations classified according to their algebraic transformation properties under the $SO(n+2)$ symmetry of the sphere $S^{n+1}$: scalar-type ($S$), vector-type ($V$) and tensor-type ($T$) gravitational perturbations. Tensor perturbations exists only in five or more spacetime dimensions ($n\geq 2$). Also, the isospectrality of the four-dimensional scalar (`polar') and vector (`axial') perturbations does not extend to higher dimensions. 

Ref.~\cite{Kodama:2003jz} obtained decoupled master variables $\Psi_s(r_*)$, with $r_*=\int dr/f$, for each of these perturbations which satisfy master equations of the form
\beq\label{master}
\lp \frac{d^2}{d r_*^2}+\omega^2-V_s\rp \Psi_s=0\qquad\quad  s=S,V,T\,.
\eeq
The effective radial potential for tensors is
\beq\label{V2}
V_{T}=\frac{n^2 f}{4r^2}\left[\lp 1+\frac{2\ell}{n}\rp^2-\frac1{n^2}+\lp 1+\frac1{n}\rp^2\lp\frac{r_0}{r}\rp^n\right]\,,
\eeq
which is the same as for a free massless scalar field propagating in this background. For vectors  it is
\beq\label{V1}
V_{V}=\frac{n^2 f}{4r^2}\left[\lp 1+\frac{2\ell}{n}\rp^2-\frac1{n^2}-3\lp 1+\frac1{n}\rp^2\lp\frac{r_0}{r}\rp^n\right]\,,
\eeq
and for scalars,
\beq\label{V0}
V_{S}=\frac{f(r)Q(r)}{4r^{2}\lp2\mu+\frac{(n+2)(n+1)}{\sR}\rp^{2}},
\eeq
%==================================%
where $\mu=(\ell+n-1)(\ell-1)$ and we abbreviate
\beq\label{sR}
\sR=\lp \frac{r}{r_0}\rp^n\,,
\eeq
which is a radial coordinate that will be useful later on. $Q(r)$ is defined as
%============<Equation>=============%
%
\begin{eqnarray}
Q(r) &=& \frac{(n+2)^2(n+1)^4}{\sR^3}\notag\\  
&&+\left( 4\mu\lp 2(n+3)^2-11(n+3)+18\rp+(n+2)(n^2-1)(n-3)\right)\frac{(n+2)(n+1)}{\sR^2}\notag\\
&&-\left( (n-3)\mu+(n+2)(n^2-1)\right)\frac{12(n+1)\mu}{\sR}\notag\\
&&+16\mu^3+4(n+3)(n+1)\mu^2\,.
\eeqa

For considering large frequencies $\omega=\mc{O}(n/r_0)$ and angular momentum numbers $\ell=\mc{O}(n)$, it will be convenient to introduce
\beq
\hat\omega=\frac{\omega}{n}\,,\qquad \hat\ell=\frac{\ell}{n}\,.
\eeq

\subsection{Decoupling and non-decoupling quasinormal modes}

Quasinormal modes are solutions of \eqref{master} characterized by the absence of any amplitudes coming in from infinity or coming out of the horizon.
Using the coordinate in \eqref{sR},
the ingoing boundary condition at the future horizon at $\sR=1$  is implemented by writing the master field as
\beq\label{horbc}
\Psi_s(\sR) = (\sR -1)^{-i\omega r_0/n}\phi_s(\sR)
\eeq
with $\phi_s(\sR)$ regular at $\sR=1$. 

\begin{figure}[t]
 \begin{center}
  \includegraphics[width=.5\textwidth,angle=0]{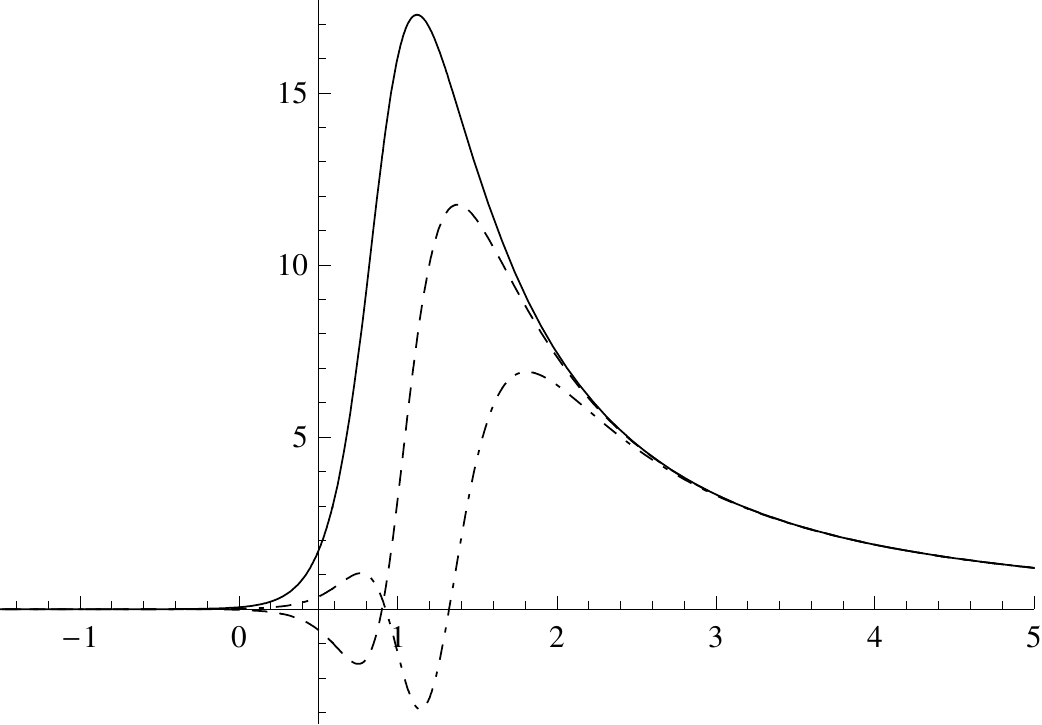}
   \end{center}
 \vspace{-5mm}
 \caption{\small Radial potential $V_s(r_*)$ for perturbations of the Schwarzschild black hole for $n=7$ and $\ell=2$. The horizon is at $r_*\to-\infty$. We use the coding solid/dashed/dot-dashed $=$ tensor/vector/scalar in this and in the next two figures. Units are $r_0=1$. }
 \label{fig:n7l2}
\end{figure}

Fig.~\ref{fig:n7l2} illustrates $V_s(r_*)$ for moderate values of $n$ and $\ell$. There is a barrier, which grows with $\ell$, corresponding to radial gradients and centrifugal energy. For small enough $\ell/n$, the scalar and vector potentials possess additional minima and maxima closer to the horizon, which are absent for the tensor perturbations.

\begin{figure}[t]
 \begin{center}
  \includegraphics[width=.47\textwidth,angle=0]{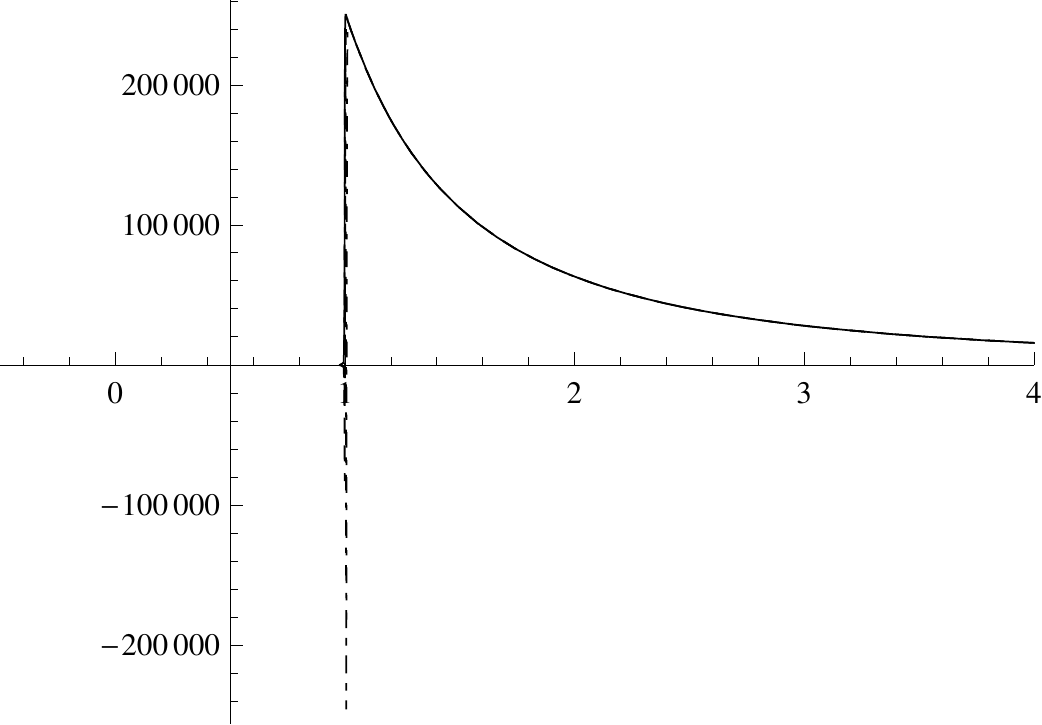}
  \hspace{5mm}
  \includegraphics[width=.47\textwidth,angle=0]{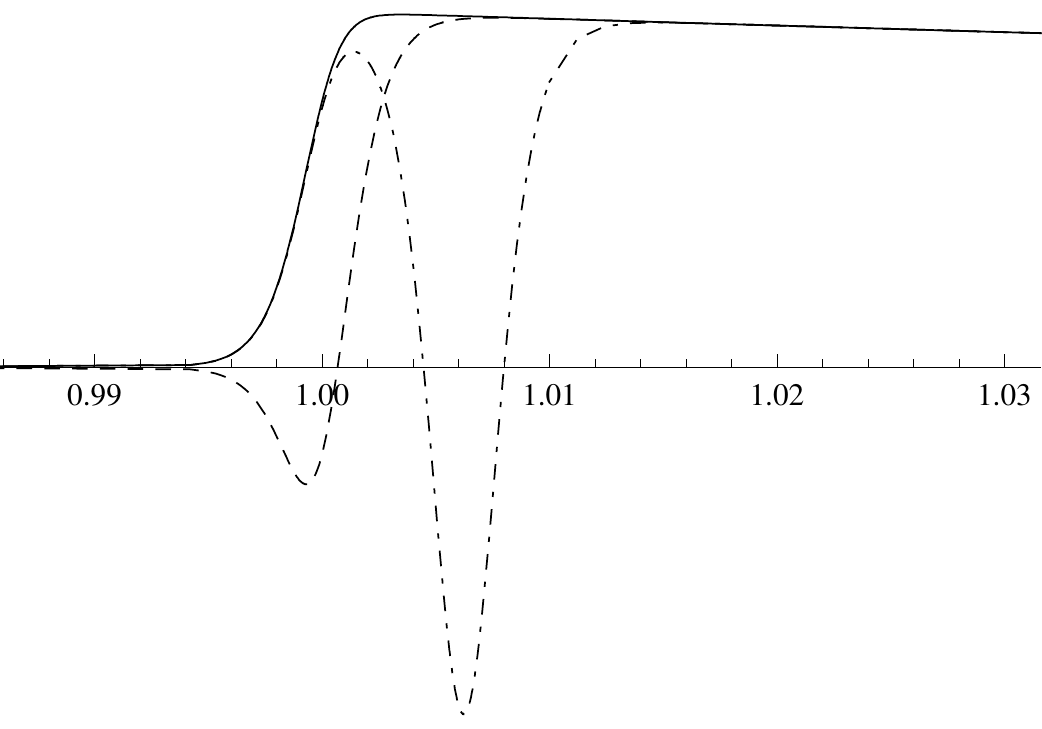}
   \end{center}
 \vspace{-5mm}
 \caption{\small Radial potential $V_s(r_*)$ for $n=1000$ and $\ell=2$. On the right is a blow-up of the potential near the peak at $r_*\simeq 1$.}
 \label{fig:n1000l2}
\end{figure}

\begin{figure}[t]
 \begin{center}
  \includegraphics[width=.47\textwidth,angle=0]{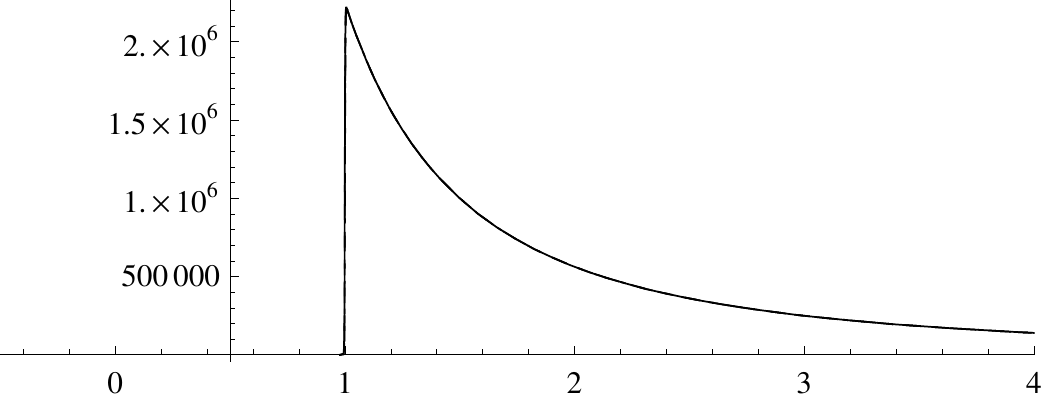}
  \hspace{5mm}
  \includegraphics[width=.47\textwidth,angle=0]{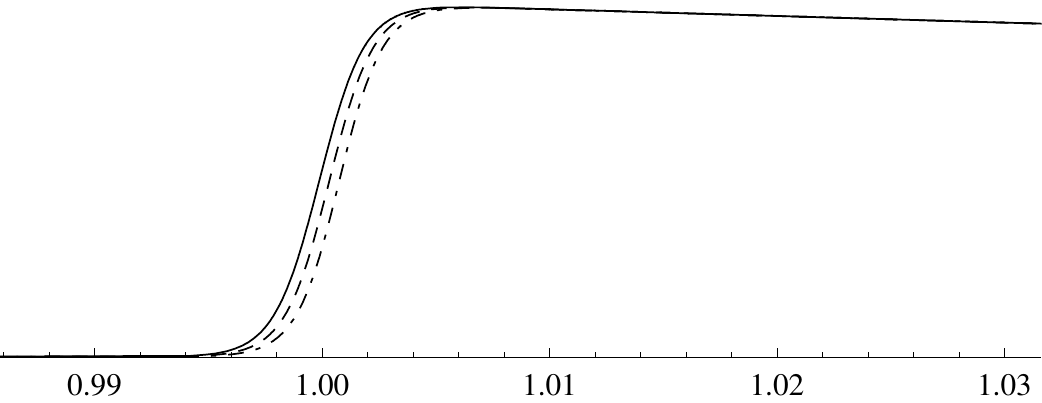}
   \end{center}
 \vspace{-5mm}
 \caption{\small Radial potential $V_s(r_*)$ for $n=1000$ and $\ell=1000$. On the right is a blow-up of the potential near the peak at $r_*\simeq 1$.}
 \label{fig:n1000l1000}
\end{figure}

Figs.~\ref{fig:n1000l2} and \ref{fig:n1000l1000} illustrate the potentials for very large $n$. We choose $n=1000$, and  $\ell=2$ and $\ell=1000$ as two representative cases of $\ell=\mc{O}(1)$  and $\ell=O(n)$. 

The height of the potentials grows like $n^2/r_0^2$, with the maximum approaching
\beq\label{Vmax}
V_s^\text{max}\to n^2\omega_c^2
\eeq
where
\beq
\omega_c=\frac1{2r_0}\lp 1+\frac{2\ell}{n}\rp\,.
\eeq
As a consequence, waves with frequency $\omega=\mc{O}(1/r_0)\ll n\omega_c$ cannot penetrate the potential: they stay either outside or inside the barrier, since their tunneling probability is infinitely suppressed as $n\to \infty$. 

We can now expect to find quasinormal modes as solutions that connect outgoing and ingoing waves by joining them below the peak of the potential, with $\text{Re}\,\hat\omega< \omega_c$. 
The potentials that these modes `see' are the ones on the left in figs.~\ref{fig:n1000l2} and \ref{fig:n1000l1000}. 
The tensor potential for all $\ell$, and the vector and scalar potentials for $\ell=\mc{O}(n)$, all approach the form
\beq\label{Vslimit}
V_s\to \frac{n^2\omega_c^2r_0^2}{r_*^2}\Theta(r_*-r_0)\,.
\eeq
Therefore the frequency spectrum will be shared by the three kinds of perturbations. For each $\ell$ there is a sequence of modes, called  `overtones', whose wavefunctions have $k-1$ nodes, $k=1,2,\dots$. The first overtones --- the least damped of these modes, with $k\ll n$ --- are sensitive only to the structure near the tip of the potential, which approaches a triangular shape that makes it easy to obtain their frequencies \cite{Emparan:2014cia}. We will also identify higher overtones that probe lower heights of the potential and have damping ratios of order one. 

But we may also seek quasinormal modes in the form of waves with frequency $\omega=\mc{O}(1/r_0)$ that are ingoing at the horizon and are trapped inside the barrier, with wavefunctions that vanish exponentially in $n$ outside the barrier, so they satisfy the condition that any incoming component is absent. These modes should be sensitive to the features of the potential in the near-horizon region, which is shown in the plots on the right in figs.~\ref{fig:n1000l2} and \ref{fig:n1000l1000}. Radial gradients in this region are large, with the derivatives scaling like $n$. Since also $V_s\propto n^2$, it follows that frequencies of order $\omega=\mc{O}(1/r_0)$ do not enter eq.~\eqref{master} to leading order in the large $n$ expansion. So these quasinormal modes, with $\hat\omega=0$, correspond to static, zero-energy states in the potential. 

Such states can only exist if the potential has a negative minimum, which occurs in the vector and scalar potentials with $\ell=\mc{O}(1)$, fig.~\ref{fig:n1000l2}~(right), but not in any other cases. 
The zero-energy states in these potentials for a given $\ell$ are unique, with no other overtones close to them. 

\subsection{Near-horizon geometry}

The near-horizon zone is conveniently described in terms of the radial coordinate \eqref{sR} as the region where $\sR\ll e^n$. The limiting geometry is
\beq
ds^2\to -\lp 1-\frac1{\sR}\rp dt^2+\frac{r_0^2}{n^2}\frac{d\sR^2}{\sR(\sR-1)}+r_0^2 d\Omega_{n+1}\,.
\eeq
The smallness in the radial direction is apparent from the prefactor $1/n^2$ in $g_{\sR\sR}$. If we separate the angular part, we obtain a finite geometry by introducing a near-horizon time coordinate $\hat t= n t$. Then fields move in the  geometry
\beq\label{nhgeom}
\frac{n^2}{r_0^2}ds^2_\text{nh}=-\lp 1-\frac1{\sR}\rp \frac{d\hat t^2}{r_0^2} +\frac{d\sR^2}{\sR(\sR-1)}\,,
\eeq
which is the two-dimensional string theory black hole of \cite{Mandal:1991tz,Elitzur:1991cb,Witten:1991yr}, as observed in \cite{Emparan:2013xia} (see also \cite{Soda:1993xc,Grumiller:2002nm}). Propagating modes in this spacetime have frequencies $\hat\omega=\mc{O}(1/r_0)$. Instead, modes with frequency $\omega=\mc{O}(1/r_0)$ are effectively static to leading order in $1/n$, and therefore, as discussed above, are obtained as zero-energy states of the potential $V_s$. 

At large $\sR$, \eqref{nhgeom} becomes the linear dilaton vacuum of string theory. The wave equations in this region have the form
\beq\label{oveq}
\lp \frac{d^2}{d(\ln\sR)^2}-\lp\omega_c^2-\hat\omega^2\rp r_0^2\rp\Psi_s=0\qquad (\sR\gg 1)
\eeq
(the metric perturbation $h_{\mu\nu}$ or a masless scalar field in \eqref{schtang} would correspond to $\Psi_s/\sqrt{\sR}$). The solutions of \eqref{oveq} are
\beq\label{oversol}
\Psi_s =  A_+\, \Psi^+(\sR)
+A_-\,\Psi^-(\sR)\,,
\eeq
where
\beq\label{psipm}
\Psi^\pm(\sR) = \sR^{\pm \sqrt{\omega_c^2-\hat\omega^2}\,r_0}\,.
\eeq
Propagating fields with real frequency $\hat\omega>\omega_c$ have non-zero flux across the asymptotic boundary at $\sR\gg 1$, and thus violate the unitarity bound, of Breitenlohner-Freedman (BF)-type, in the near-horizon spacetime.\footnote{For static states of a two-dimensional massive scalar field in \eqref{nhgeom}, the BF bound on the mass is $ m^2\geq -1/4$~\cite{Emparan:2013oza}. The two-dimensional mass of $\hat\ell$-waves is $m^2_\ell=\omega_c^2-1/4 =\hat\ell(\hat\ell+1)\geq 0$, \ie\ above the bound.}

When $\text{Re}\,\hat\omega<\omega_c$, which implies that $\text{Re}\,\sqrt{\omega_c^2-\hat\omega^2}>0$, the solutions that approach $\Psi^-$ at large $\sR$ are normalizable, while $\Psi^+$ is non-normalizable.

States with $\hat\omega=\omega_c$ are at the BF bound, and their general form at $\sR\gg 1$ is
\beq\label{ABform}
\Psi \sim A+B\ln\sR\,.
\eeq 
These are not normalizable.

\bigskip

The method to solve the equations exploits the separation of scales $r_0/n\ll r_0$ to perform a matched asymptotic construction, matching the near- and far-zone solutions over the region $r_0/n\ll r-r_0\ll r_0$ where they overlap.
The overlap-zone is the asymptopia of the near-zone, so in terms of $\sR$ it is defined as
\beq
1\ll\sR\ll e^n\,.
\eeq

\section{Far- and near-zone solutions}
\label{sec:farnear}

\subsection{Far-zone solutions}

In the far-zone where $r-r_0\gg r_0/n$, the terms $(r_0/r)^n$ are exponentially small in $n$. Thus we can set $f=1$ in the wave equation, which then becomes the same as in Minkowski spacetime. Up to normalization, the outgoing waves are Hankel functions,
\beq\label{outmode}
\Psi_s=\sqrt{r}\,H^{(1)}_{n\omega_c r_0}(\omega r)\,.
\eeq
The radial dependence of the metric perturbation is $h_{\mu\nu}\sim r^{-(n+1)/2}\Psi_s$. 

This solution can now be taken to the overlap zone $r-r_0\ll r_0$. In terms of the coordinate $\sR$ it takes the form \eqref{oversol} with coefficients $A_\pm(\omega)$ computed in  \cite{Emparan:2013moa} and app.~\ref{app:debye}, and whose relevant properties will be discussed below. The expansion \eqref{oversol} is valid only if $\hat\omega$ differs from $\omega_c$ by more  than $\mc{O}\lp n^{-2/3}\rp$. Let us discuss the three relevant cases.

\subsubsection{$\hat\omega>\omega_c$: above the BF bound} In this case $A_-(\omega)=0$ \cite{Emparan:2013moa}. The solution is oscillating and the wave remains purely outgoing in the overlap region: it travels above the peak of the potential and transmission is perfect. Since in the near-horizon region there is not any other higher peak to scatter it back, the wave must remain outgoing also at the horizon. Hence it is impossible to satisfy the ingoing boundary condition there, and there are no quasinormal modes with these frequencies. 

States with $\hat\omega>\omega_c$ violate the BF bound on scalars in the geometry \eqref{nhgeom}. The violation of unitarity corresponds to the states being able to freely leave or enter the near-horizon region.

\subsubsection{$|\hat\omega|^2<\omega_c^2$} 

At these frequencies we find (see app.~\ref{app:debye})
\beq\label{A+A-far}
\left| \frac{A_+(\omega)}{A_-(\omega)}\right|_\text{far}= e^{-2 n\omega_c r_0 \text{Re}\,f\lp\hat\omega/\omega_c\rp}
\eeq
with
\beq\label{fz}
f(z)=\ln\lp\frac{1+\sqrt{1-z^2}}{z}\rp-\sqrt{1-z^2}\,.
\eeq
The function $\text{Re}\,f(z)$ is non-zero on the complex $z$ plane except on a line (to be discussed below), which implies that in general one of the two amplitudes is suppressed exponentially in $n$ relative to the other one, so it is too small to be obtained in a near-horizon analysis in a power-series expansion in $1/n$. For instance, if $\text{Im}\,\hat\omega/\omega_c$ is small enough then the non-normalizable component is suppressed.

\subsubsection{$\omega_c-\hat\omega=\mc{O}\lp n^{-2/3}\rp$} 
\label{sec:homomc}

We may have waves with frequency just below the peak of the potential for which the transmitted amplitude is not suppressed factorially in $n$. Take
\beq\label{omqnm}
\hat{\omega} = \omega_{c} -\left(\frac{e^{i\pi}\omega_{c}}{2n^{2}r_0^2}\right)^{1/3}\delta \omega\,,
\eeq
with $\delta\omega=\mc{O}(1)$, and where the prefactors have been chosen for later convenience. For these frequencies the expansion that gives \eqref{A+A-far} is not valid and instead one gets
\beq\label{aiexpand}
\Psi_{s} \propto \text{Ai}(-\delta\omega) +\text{Ai}'(-\delta\omega)\left(\frac{2\omega^{2}_{c}}{ne^{i\pi}}\right)^{1/3}\ln{\sR}+O(n^{-2/3})\,, 
\eeq
%
%==================================%
where Ai is the Airy function. This is of the form \eqref{ABform}, with
\beq
\label{ABfar}
\left| \frac{A}{B}\right|_\text{far}\sim n^{1/3}\,. 
\eeq

\bigskip

One situation of potential relevance that we are not covering here is when $\text{Re}\,\hat\omega<\omega_c$ but $|\hat\omega|^2>\omega_c^2$. Our approximations do not apply in this regime, in which the modes are very strongly damped. We will make some more comments in sec.~\ref{sec:omegaplane}.

\subsection{Near-zone solutions}
\label{sec:nearsols}

The equations \eqref{master} for tensors and vectors, to leading order in $1/n$ in the near-zone, are of hypergeometric type. The solutions that satisfy the horizon boundary condition in the form \eqref{horbc} are, for tensors \cite{Emparan:2013moa},
\beq\label{nhs2}
\Psi_{T} (\sR)=(\sR -1)^{-i\hat\omega r_0}\,\sqrt{\sR}\,{}_2F_1\lp q_+,q_-,q_++q_-;\sR-1\rp\,,
\eeq
and for vectors,
\beq\label{nhs1}
\Psi_{V} (\sR)=(\sR -1)^{-i\hat\omega r_0}\,\sR^{3/2}\,{}_2F_1\lp 1+q_+,1+q_-,1+q_++q_-;\sR-1\rp\,,
\eeq
where
\beq
q_\pm=\frac12-i\hat\omega r_0\pm \sqrt{\omega_c^2-\hat\omega^2}\,r_0\,.
\eeq
For scalar-type perturbations the equation is more complicated, but in appendix~\ref{app:scOn} we show that the appropriate solution is 
\beq\label{nhs0}
\Psi_{S} (\sR) = (\sR -1)^{-i\hat\omega r_0}\frac{\sqrt{\sR}}{1+2\hat\ell(\hat\ell+1) \sR}\, \cD_1\  {}_2 F_1(1-q_+,1-q_-,2-q_+-q_-;1-\sR)\,,
\eeq
where $\cD_1$ is the differential operator defined in \eqref{cD1}.

These solutions may look complicated, but the only information that we need from them is their asymptotic behavior in the overlap zone where $\sR\gg 1$. The case $\hat\omega= 0=\hat\ell$ is special and we will discuss it separately in sec.~\ref{sec:decspec}. For other generic $\hat\omega$ and $\omega_c$, it is easy to find that at large $\sR$ these solutions contain both the normalizable and non-normalizable components $\Psi^{\pm}$ with amplitudes of the same order in $n$,\footnote{When $2\sqrt{\omega_c^2-\hat\omega^2}r_0\in \N$ there appear subleading terms $\sim (\ln\sR)/\sR$ in the normalizable wavefunction but the amplitudes satisfy the ratio \eqref{A+A-near}.}
\beq\label{A+A-near}
\left| \frac{A_+(\hat\omega)}{A_-(\hat\omega)}\right|_\text{near}=\mc{O} \lp 1\rp\,.
\eeq
The large $\sR$ expansion is different when $\hat\omega=\omega_c$, in which case we obtain \eqref{ABform} with 
\beq\label{ABnear}
\left| \frac{A}{B}\right|_\text{near}=\mc{O} \lp 1\rp\,. 
\eeq
We will not need the detailed values of these ratios, but only the fact that the two amplitudes are of the same order in $n$. Actually, we should expect that a horizon boundary condition generically results in comparable amplitudes of the two independent components at $\sR\gg 1$. This is one of the main assumptions that underlie the universality of the result in \cite{Emparan:2014cia}. We will return to it in sec.~\ref{sec:univer}.

\medskip

The matching of far- and near-zone solutions is only possible for specific values of the frequency. There are two different ways to achieve this, which lead to two different sets of quasinormal modes.

\section{Non-decoupling modes}
\label{sec:nondec}

This class of modes is obtained by considering frequencies for which the generic near-horizon conditions \eqref{A+A-near} and \eqref{ABnear} hold --- so the modes are non-normalizable ---, which restricts the frequencies of far-zone outgoing waves.

\subsection{Least-damped modes}%: $\omega_c-\hat\omega = \mc{O}(n^{-2/3})>0$}

The far-zone result \eqref{ABfar} is incompatible for general $\delta\omega$ with the near-zone one \eqref{ABnear}. But we can match the solutions if we require that\footnote{$n$-independent rescalings of $\sR$ are allowed that can generate a constant term of the same order as the $\ln\sR$ term.} 
%============<Equation>=============%
%
\begin{eqnarray}
\text{Ai}(-\delta\omega) = 0\,,
\end{eqnarray}
%
%==================================%
\ie\ quasinormal frequencies are in correspondence with the zeroes of the Airy function. These all occur at negative values of the argument, $\text{Ai}(-a_k)=0$, so
\beqa
\delta\omega&=&a_k\notag\\
&\simeq& \lp\frac{3\pi}{8}(4k-1)\rp^{2/3}\,,
\eeqa
with $k=1,2,\dots$. The second line is the asymptotic approximation to the Airy zeroes, which is very accurate (to better than $1\%$) even for $a_1$.
From this result and \eqref{omqnm} we find the quasinormal frequency spectrum
%============<Equation>=============%
%
\beq\label{onomega}
\omega r_0 = \frac{n}2+\ell -a_{k}\left(\frac{e^{i\pi}}{2}\lp \frac{n}2+\ell\rp\right)^{1/3}\,.
\eeq
%
%==================================% 
Equivalently,
\beq\label{Reonomega}
\text{Re}\,\omega r_0 = \frac{n}2+\ell -\frac{a_{k}}{2^{4/3}}\lp \frac{n}2+\ell\rp^{1/3}\,,
\eeq
and 
\beq\label{Imonomega}
\text{Im}\,\omega r_0 = -\frac{\sqrt{3}\,a_{k}}{2^{4/3}}\lp \frac{n}2+\ell\rp^{1/3}.
\eeq
The real part of the frequency is slightly below $n\omega_c$, as expected, and the
imaginary part is negative, in accord with the stability of the Schwarzschild 
black hole \cite{Ishibashi:2003ap}.

The index $k$ corresponds to the number $k-1$ of nodes of the perturbation and labels different overtones of the quasinormal modes for a given $\ell$. Higher overtones have lower $\text{Re}\,\omega$ and higher $|\text{Im}\,\omega|$ \ie\ lower overtones are less damped. Our approximations break down when $k\sim n$. 

The damping ratio of these modes
\beq
\frac{\text{Im}\,\omega}{\text{Re}\,\omega}\sim n^{-2/3}
\eeq
vanishes as $n\to\infty$ and thus these modes are long-lived in their characteristic time scale. They limit to undamped normal modes.

In ref.~\cite{Emparan:2014cia} we exploited the fact that the potential near its maximum takes a triangular shape in order to give a simple, universal derivation of the spectrum of quasinormal frequencies in this range. Our more detailed derivation here demonstrates that complete explicit solutions can be found which satisfy the required boundary conditions.

\subsection{Higher overtones}%: $\textnormal{Re}\,\hat\omega<\omega_c$ or $|\hat\omega|^2<\omega_c^2$ or $\hat\omega^2<\omega_c^2$ or $\textnormal{Re}\,\sqrt{\omega_c^2-\hat\omega^2}>0$}

When $|\hat\omega|^2<\omega_c^2$ we find again that \eqref{A+A-far} and \eqref{A+A-near} are incompatible except if the exponent in  \eqref{A+A-far} is $\mc{O}(1)$, which requires that 
\beq\label{hiover}
\text{Re}\,f(\hat\omega/\omega_c)=0\,,
\eeq
with $f$ given in \eqref{fz}. This equation determines a set of quasinormal frequencies to leading order in $1/n$. It gives a continuous spectrum, which should be discretized into separate overtones when one includes the next correction in the large $D$ expansion (which does not seem easy to obtain).
The equation is transcendental and does not admit any simple explicit form, but nevertheless it is easily solved numerically and we plot it in fig.~\ref{fig:ODmodes}. 
\begin{figure}[t]
 \begin{center}
  \includegraphics[width=.6\textwidth,angle=0]{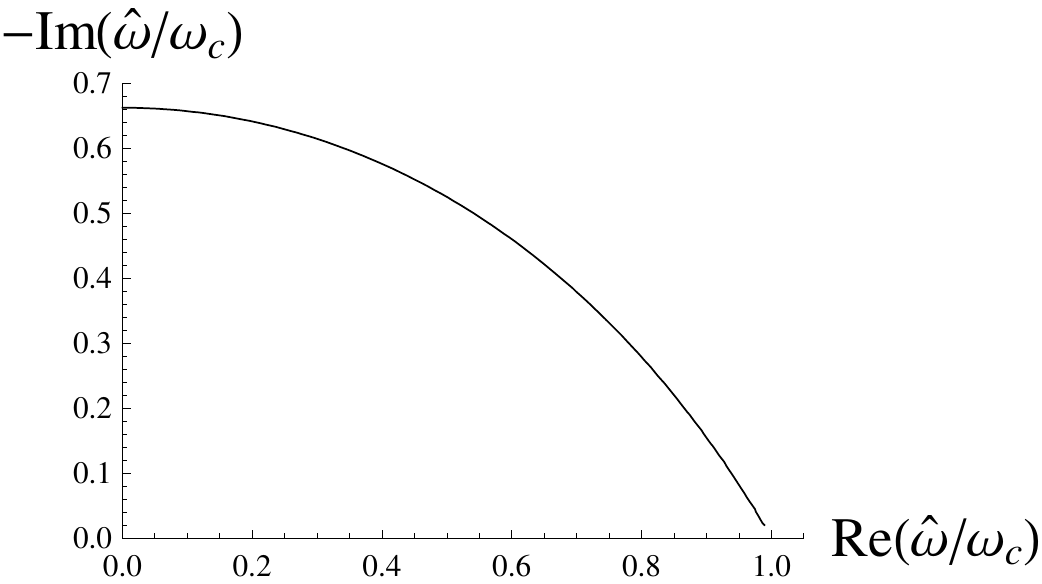}
   \end{center}
 \vspace{-5mm}
 \caption{\small Solution of \eqref{hiover} determining quasinormal frequencies for $0<|\hat\omega|^2<\omega_c^2$. The continuous line of frequencies should become a discrete spectrum when higher order terms at large $n$ are included. Near $\hat\omega=\omega_c$ the curve connects smoothly to the spectrum \eqref{onomega}.}
 \label{fig:ODmodes}
\end{figure}

The curve reaches the imaginary axis at $-\text{Im}\,(\hat\omega/\omega_c)\simeq 0.6627$. For frequencies close to the real axis we find that 
\beq\label{ReomImom}
\text{Re}\,\hat\omega \simeq \omega_c+ \frac1{\sqrt{3}}\,\text{Im}\,\hat\omega\,.
\eeq
Since in this region $\hat\omega\to\omega_c$, the spectrum should be replaced by  \eqref{Reonomega}, \eqref{Imonomega}. In fact, the latter also has the form \eqref{ReomImom}, so it continues smoothly into the higher-overtone regime given by \eqref{hiover}. 

For these modes both $\text{Re}\,\omega$ and $\text{Im}\,\omega$ are $\mc{O}(n/r_0)$, so their damping ratio is 
\beq
\frac{\text{Im}\,\omega}{\text{Re}\,\omega}=\mc{O}(1)\,.
\eeq

\subsection{Universality of non-decoupling spectrum}
\label{sec:univer}

This analysis has not required any detailed information about the near-horizon solutions, only the generic results \eqref{A+A-near} and \eqref{ABnear} for the amplitude ratios, which then constrain the far-zone waves. But the latter are actually waves in flat spacetime, which effectively propagate in a potential \eqref{Vslimit} that is abruptly cut off at $r=r_0$.
This was argued in \cite{Emparan:2013moa} to be the universal far-zone structure for all static, spherically symmetric black holes in the limit $D\to\infty$. 

This implies that the non-decoupling spectra \eqref{onomega} and \eqref{hiover} are universally present in all asymptotically flat, static, spherically symmetric black holes in the limit $D\to\infty$.\footnote{If the asymptotic conditions change, like in AdS, these may not be quasinormal modes, although their frequencies can still play a role in the relaxation of the black hole.} This extends the findings of \cite{Emparan:2014cia} to the spectrum given by  \eqref{hiover}. In appendix~\ref{app:extr} we show that the result also applies to extremal charged black holes, in an illustrative case where the equations can be solved explicitly, thus resolving a potential issue mentioned in \cite{Emparan:2014cia}.

Finally, note that the additional extrema of the scalar and vector potentials for $\ell=\mc{O}(1)$ are not expected to modify the spectrum to leading large $n$ order, since in the limit $n\to\infty$ the positions in $r_*$ of these maxima and minima is coincident. So the universal spectrum applies both for $\ell=\mc{O}(n)$ and $\ell=\mc{O}(1)$, but in the latter case the accuracy of the result at finite $n$, in particular for scalar modes, may be affected.

\section{Decoupled modes}
\label{sec:decspec}

Here we seek near-horizon solutions at special frequencies such that their large $\sR$ behavior is not the generic one in \eqref{A+A-near}, but instead has vanishing non-normalizable component, $A_+=0$, so
\beq\label{nhbc}
\Psi_s\propto \Psi^{-}(\sR)\,.
\eeq
In contrast to the previous sector of non-decoupling modes, these normalizable modes involve detailed properties of the wave equation in the near-horizon region.
Our analysis in sec.~\ref{sec:nearsols} leaves only one possibility for \eqref{nhbc}, namely modes with $\hat\omega=0=\hat\ell$, \ie\ with $\omega=\mc{O}(1/r_0)$ and $\ell=O(1)$.
Any normalizable solutions in this frequency range are static to leading order, so one needs to go to higher orders in $1/n$ to determine the quasinormal frequencies.\footnote{While it is clear that it is possible to match these modes to an outgoing far-zone wave, in order to distinguish between far-zone outgoing and ingoing waves, one needs to include a near-zone non-normalizable amplitude that is exponentially small in $n$, \ie\ cannot be obtained at any order in $1/n$ perturbation theory.}

The method to solve the equations is conventional in perturbation analysis. Beginning from the exact equation in the form
\beq
\lp\mc{L}+U_s\rp\Psi_s(\sR)=0
\eeq
where
\beq
\mc{L}\Psi=-\frac{\sR-1}{\sR^{1/n}}\frac{d}{d\sR}\lp\frac{\sR-1}{\sR^{1/n}}\frac{d}{d\sR}\Psi\rp
\eeq
and 
\beq
U_s=\frac1{n^2}\lp V_s(\sR)-\omega^2\rp\,,
\eeq
we expand all quantities in powers of $1/n$ as
\beq
\Psi_s=\sum_{k\geq 0}\frac{\Psi_s^{(k)}}{n^k}\,,\qquad
\mc{L}=\sum_{k\geq 0}\frac{\mc{L}^{(k)}}{n^k}\,,\qquad
U_s=\sum_{k\geq 0}\frac{U_s^{(k)}}{n^k}\,,\qquad
\omega=\sum_{k\geq 0}\frac{\omega_{(k)}}{n^k}\,.
\eeq
The equations become of the form
\beq
\lp\mc{L}^{(0)}+U_s^{(0)}\rp\Psi_s^{(k)}=\mc{S}^{(k)}
\eeq
where
\beq
\mc{L}^{(0)}\Psi=-(\sR-1)\frac{d}{d\sR}\lp(\sR-1)\frac{d}{d\sR}\Psi\rp
\eeq
and for $k\geq 1$ the source terms $\mc{S}^{(k)}$ are obtained from  $\mc{L}^{(j)}+U_s^{(j)}$ with $j\leq k$, and from the solutions $\Psi_s^{(j)}$ with $j<k$. If we have the two independent solutions $u_0(\sR)$, $v_0(\sR)$ to the leading order homogeneous equation, then the successive solutions can be obtained perturbatively by the method of variation of constants.

The boundary condition at $\sR\gg 1$ \eqref{nhbc} is
\beq\label{asybc}
\Psi(\sR\to\infty)\to \frac1{\sqrt{\sR}}\,,
\eeq
\ie\ the non-normalizable terms $\sim\sqrt{\sR}$ must be absent.
This is the same condition at all orders in the expansion in $1/n$,\footnote{At higher orders there can be terms $\ln\sR/n$ multiplying $1/\sqrt{\sR}$. These are allowed since in this region we assume $\sR\ll e^n$. We will find them below.} so the mode remains normalizable --- hence decoupled --- at all perturbative orders in $1/n$.
Regularity at the future horizon \eqref{horbc} gives different conditions at each order,
\beqa\label{hborders}
\Psi^{(0)}(\sR\to 1)&\to& 1\label{hbc0}\,,\\
\Psi^{(1)}(\sR\to 1)&\to& -i\omega_{(0)} \ln(\sR-1)\,,\label{hbc1}\\
\Psi^{(2)}(\sR\to 1)&\to& -i\omega_{(1)} \ln(\sR-1)-\frac12\omega_{(0)}^2\lp\ln(\sR-1)\rp^2\,,
\eeqa
etc., where we have (arbitrarily) fixed the overall amplitude factor and have set, also for the remainder of this section,
\beq
r_0=1\,.
\eeq

Since the procedure is straightforward, we only give details of the calculation of the leading-order frequencies.

\subsection{Tensor-type modes}

The tensor potential \eqref{V2} gives
\beq
U_{T}=\frac{\sR-1}{4\sR^{1+2/n}}\left[\lp 1+\frac{2\ell}{n}\rp^2-\frac1{n^2}+\frac{1}{\sR}\lp 1+\frac1{n}\rp^2 \right]-\frac{\omega^2}{n^2}\,,
\eeq
so
\beq\label{UT0}
U_{T}^{(0)}=\frac{\sR^2-1}{4\sR^2}\,,
\eeq
and the leading order independent solutions are
\beq
u_0=\sqrt{\sR}\,,\qquad v_0=\sqrt{\sR}\,\ln\lp 1-\sR^{-1}\rp\,.
\eeq
The two boundary conditions \eqref{asybc} and \eqref{hbc0}
are impossible to satisfy simultaneously, so there are no decoupled quasinormal modes of tensor type. This was indeed expected given the absence of minima in the potential $V_T$.

\subsection{Vector-type modes}

The vector potential \eqref{V1} gives
\beq
U_{V}=\frac{\sR-1}{4\sR^{1+2/n}}\left[\lp 1+\frac{2\ell}{n}\rp^2-\frac1{n^2}-\frac{3}{\sR}\lp 1+\frac1{n}\rp^2 \right]-\frac{\omega^2}{n^2}\,,
\eeq
so\footnote{Changing to $\hat r_*=\ln(\sR-1)$ in $U_{V}^{(0)}$ reproduces the form of the vector potential in  fig.~\ref{fig:n1000l2}~(right).}
\beq
U_{V}^{(0)}=\frac{(\sR-1)(\sR-3)}{4\sR^2}\,,
\eeq
and the leading order independent solutions are
\beq
u_0=\frac1{\sqrt{\sR}}\,,\qquad v_0=\frac{\sR+\ln\lp \sR-1\rp}{\sqrt{\sR}}\,.
\eeq
The boundary conditions \eqref{asybc} and \eqref{hbc0} select
\beq
\Psi_{V}^{(0)}= u_0\,.
\eeq
So there does exist a vector quasinormal mode, although its frequency, as explained before, is not determined at this order.

At the next order, the solution that satisfies \eqref{asybc} is found to be
\beq
\Psi_{V}^{(1)}=A_1 u_0-\frac{(\ell-1)\ln(\sR-1)+\ln\sqrt{\sR}}{\sqrt{\sR}}\,,
\eeq
with integration constant $A_1$.
The boundary condition at the horizon \eqref{hbc1} 
selects $A_1=0$ and determines the frequency as
\beq
\omega_{(0)}=-i(\ell-1)\,.
\eeq
Observe that the frequency is determined by the horizon boundary condition, and not through its appearance in the equation via $U_{V}$, which is at a higher order in the expansion. This feature recurs through all higher orders in the perturbation analysis.

It is  straightforward to carry the calculation to higher orders, the limit being the ability to perform the successive integrations in analytic form. We have done them up to $1/n^3$, finding
\beqa\label{vqnm}
\omega&=&-i(\ell-1)\Biggl( 1+\frac1{n}(\ell-1)+\frac{2}{n^2}(\ell-1)\lp\frac{\pi^2}{6}-1\rp\notag\\
&&\qquad\qquad\quad
+\frac{4}{n^3}(\ell-1)\lp 1-\ell\zeta(3)+(\ell-1)\frac{\pi^2}{6}\rp\Biggr)\,,
\eeqa
or, perhaps more suggestively,
\beqa
\omega&=&-i(\ell-1)\Biggl( 1+\lp 1+\frac{2\lp\zeta(2)-1\rp}{n} - \frac{4\lp\zeta(3)-1\rp}{n^2}\rp\frac{\ell-1}{n}
\notag\\
&&\qquad\qquad\quad
+\frac{4\lp\zeta(2)-\zeta(3)\rp}{n^3}(\ell-1)^2\Biggr)\,.
\eeqa

Notice that the modes are purely imaginary, and that for a given value of $\ell$ they are unique, so they are isolated in the complex $\omega$ plane without any other overtones nearby them.

\subsection{Scalar-type modes}

For the scalar modes a technical complication arises when using the formulation of the problem in the master-variable form \eqref{master} of \cite{Kodama:2003jz}. 
If one takes the large $n$ limit of the scalar potential $V_{S}(\sR)$, the leading order term is \eqref{UT0}, the same as for tensor perturbations. Since we have found that this potential does not admit  normalizable zero energy states, naively one would conclude that there cannot be any scalar quasinormal modes. However, this limit misses the presence of the non-trivial maxima and minima of the scalar potential in the near-horizon region, which lie at $\sR \sim n$, \ie\ still within the near-zone $\sR \ll e^n$. 
This is problematic, since the denominator in $V_{S}$ in \eqref{V0} (introduced through the definition of $\Psi_S(\sR)$ in \cite{Kodama:2003jz}) has a behavior at large $n$ that differs depending on whether $\sR=\mc{O}(1)$ or  $\sR=\mc{O}(n)$. This modifies the asymptotic behavior of the solution, even at the leading order in the expansion. 
In order to properly deal with the region where $\sR=\mc{O}(n)$ we introduce a new variable
\beq
\bar\sR=\frac{\sR}{n}
\eeq
that remains finite in the region of interest, and
expand and solve the equations while keeping $\bar\sR=\mc{O}(1)$. These solutions can be matched at small $\bar\sR$ to those at $\sR=\mc{O}(1)$ in the new overlap zone $1\ll \sR\ll n$. 

When $\bar\sR=\mc{O}(1)$ the potential to leading order becomes
\beq
V_{S}(\bar\sR)\to n^2 \bar V_S(\bar\sR)=\frac{n^2}{4} \frac{4(\ell-1)^2\bar\sR^2-12(\ell-1)\bar\sR+1}{\lp 2(\ell-1)\bar\sR+1\rp^2}\,.
\eeq
This potential reaches two maxima of equal height $V^\text{max}_S=n^2/4$ at its endpoints, one at small $\bar\sR=\sR/n$ where it can be matched to the potential \eqref{UT0} obtained in the region $1\ll\sR\ll n$, and the other at $\bar\sR\gg 1$ where it joins the maximum \eqref{Vmax} from the far-zone. In between them, it reaches a minimum at $\bar\sR=2/(3\ell-2)$. In this way we reproduce all the features of the scalar potential in fig.~\ref{fig:n1000l2}~(right).

The leading order, homogeneous equation in this region is now
\beq
\bar\sR\, \Psi''(\bar\sR )+\Psi'(\bar\sR )-\bar V_S(\bar\sR)\Psi(\bar\sR )=0\,,
\eeq
which is solved by
\beqa
\bar u_0&=&\frac{\sqrt{\bar\sR}}{1+2(\ell-1)\bar\sR}\,,\nonum
\bar v_0&=&\frac{\sqrt{\bar\sR}}{1+2(\ell-1)\bar\sR}\lp 4(\ell-1)^2\bar\sR +4(\ell-1)\ln\bar\sR-\frac1{\bar\sR}\rp\,.
\eeqa
At $\bar\sR\to\infty$ we find that $\bar u_0\to 1/\sqrt{\bar\sR}$ and therefore satisfies the asymptotic boundary condition. On the other hand, at small $\bar\sR=\sR/n$ we find
\beq
\bar u_0 \to \sqrt{\sR/n}\,,
\eeq
which can be matched to the solution $u_0=\sqrt{\sR}$, valid where $\sR=\mc{O}(1)$ and which satisfies the boundary condition at the horizon. 

So with this new matched asymptotic expansion, entirely within the near-horizon region $\sR\ll e^n$, we have obtained a quasinormal mode solution. Again, at this order the frequency is not determined, but in appendix~\ref{app:scmaster} we explain how to iterate the calculation to the next order to find two modes, related by $\omega_-=-\omega_+^*$, with frequencies
\beq\label{omsc0}
\omega_{(0)\pm}=\pm\sqrt{\ell-1}-i (\ell-1)\,.
\eeq

The formulation of the scalar perturbations in \cite{Kodama:2003jz} using three gauge-invariant variables $X(\sR)$, $Y(\sR)$, $Z(\sR)$, instead of the single master variable $\Psi_S(\sR)$, does not change qualitatively when one considers $\sR$ of order $n$ and so does not require this second matching. It is a more practical method that we have carried through up to $1/n^3$. The details are still cumbersome, so we postpone them to appendix~\ref{app:schigher} and quote only the final result,
\begin{eqnarray}\label{sqnmRe}
 \text{Re}\,\omega_\pm &=&\pm\sqrt{\ell-1}\Biggl(1+ \frac{1}{n}\left(\frac{3 \ell}{2}-2\right)+ \frac{1}{n^2}\left(\frac{7 \ell^2}{8}+\frac{2 \pi ^2 \ell}{3}-\frac{9\ell}{2}-\frac{2\pi^2}{3} +4\right)\notag\\
&&\qquad\qquad\quad+\frac{1}{n^3}\biggl(-\frac{5 \ell^3}{16} -\ell^2\left(6 \zeta (3)+\frac{5}{2}-\frac{5 \pi ^2}{3}\right)
	\notag\\
&&\qquad\qquad\qquad\qquad-\frac{\ell}{6} 
   \lp 26 \pi ^2-72 \zeta (3)-63 \rp
-8 \zeta (3) + \frac{8 \pi^2}{3}-8\biggr)\Biggr)\,.
\end{eqnarray}
\begin{eqnarray}\label{sqnmIm}
 \text{Im}\,\omega_\pm &=&-i (\ell-1)\Biggl(1+ \frac{1}{n}(\ell-2)
+ \frac{1}{n^2}\lp 4-3\ell+(\ell-2)\frac{\pi^2}{3}\rp\notag\\
&&\qquad\qquad\quad+\frac{1}{n^3}\biggl(2 \ell^2 \left(\frac{\pi^2}{3}-2 \zeta
   (3)\right)+\ell \left(8 \zeta (3)+7-3 \pi ^2\right)
\notag\\
&&\qquad\qquad\qquad\qquad
+8 \left(\frac{\pi^2}{3}-\zeta
   (3)-1\right)\biggr)\Biggr)\,.
\end{eqnarray}
Again, there are no other overtones nearby these modes in the complex $\omega$ plane.

\section{Numerical accuracy}\label{sec:numbers}

Ref.~\cite{Dias:2014eua} contains numerical results of high precision for quasinormal frequencies up to very large values of $n$. Moreover, these values are computed not only at integer $n$ but also at fractional values separated by small steps, which can be compared with our calculations.

\subsection{Non-decoupling modes}

The results of \cite{Dias:2014eua} in this sector only allow to make comparisons to modes with low overtone number.
\begin{figure}[t]
 \begin{center}
  \includegraphics[width=.47\textwidth,angle=0]{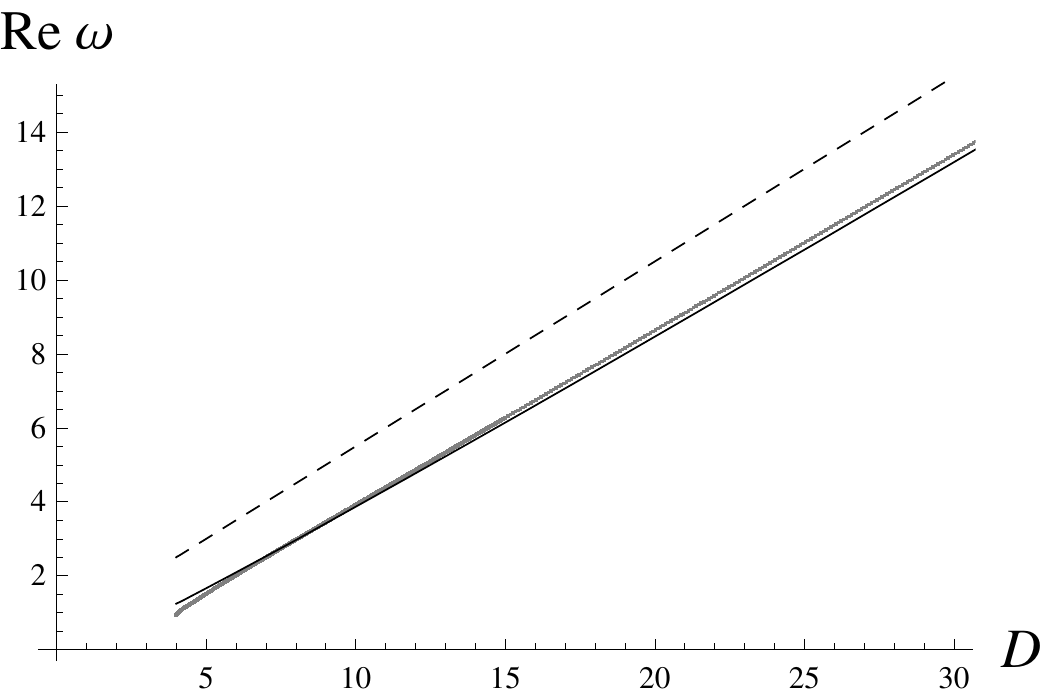}
  \hspace{5mm}
  \includegraphics[width=.47\textwidth,angle=0]{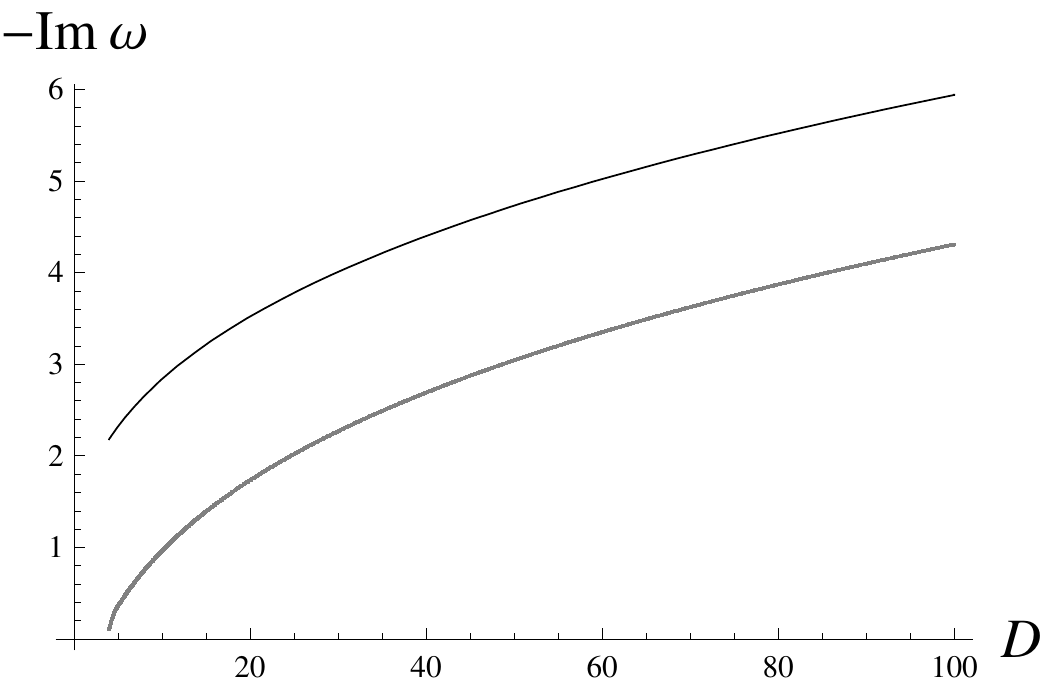}
   \end{center}
 \vspace{-5mm}
 \caption{\small Frequency of $\ell=2$, $k=0$ (fundamental) tensor quasinormal mode as a function of $D$. Solid lines: analytical results eq.~\eqref{Reonomega}, \eqref{Imonomega}; dashed line: leading order result $\omega=(D-3)\omega_c$. Gray lines: numerical results \cite{Dias:2014eua}. For $\text{Re}~\omega$ we only include data up to $D=30$ for greater clarity.}
 \label{fig:Tl1}
\end{figure}
In fig.~\ref{fig:Tl1} we compare them to \eqref{onomega}, with $n=D-3$. The real part of the frequency agrees well, to accuracy $\approx 1/(2(D-3))$. 
However, the imaginary part shows poorer agreement, with significant discrepancies even at the largest value $D=100$ computed in \cite{Dias:2014eua}. Furthermore, ref.~\cite{Dias:2014eua} found that $\text{Im}~\omega\sim D^{1/2}$ rather than $\sim D^{1/3}$ as implied by \eqref{Imonomega}. We can argue that this disagreement is not unexpected.
The $\sim D^{1/2}$ behavior is actually the one predicted by the WKB method \cite{Iyer:1986np,Konoplya:2003ii,Berti:2009kk}, which approximates the potential near its maximum by an inverted parabola. It is clear from our analysis in sec.~\ref{sec:qualia} (\eg\ figs.~\ref{fig:n1000l2} and \ref{fig:n1000l1000}), that at sufficiently large $D$ the inverted parabola must become a bad approximation. The breakdown of the WKB approximation can be estimated to occur at the value of $D$ where the
successive WKB corrections become as large as the leading result. At large $n$ one has
\beq
\left.\frac{d^j V}{dr_*^j}\right|_\text{max}\simeq -(-2)^{j-2}n^{j+1}\qquad (j\geq 2)\,,
\eeq 
which when plugged into the WKB expansion in \cite{Iyer:1986np,Konoplya:2003ii} can be seen to imply that it breaks down at $D\approx 60$. Departures of WKB from the numerical results are well apparent at around this value of $D$. Still, this does not necessarily imply that the approximation by a triangular potential becomes valid at this value of $D$. Direct inspection shows that the peak of the potential remains quite smooth, and thus \eqref{Imonomega} is not a good approximation, until around $D\sim 300$, which is higher than numerically computed, and also than what may be interesting for practical applications. 

It is remarkable, however, that  $\text{Re}\,\omega$ in \eqref{Reonomega} is significantly improved by the $\sim D^{1/3}$ correction term even at low $D$, see fig.~\ref{fig:Tl1}~(left).\footnote{This may be partly due to the fact that the relative size of the correction to the real part is $\sim D^{-2/3}$, while for the imaginary part it will be (once it is computed) $\sim D^{-1/3}$, and therefore larger.}
This phenomenon was also observed in \cite{Emparan:2013oza}, and suggests that in some respects large $D$ universal behavior may begin to become apparent at lower values of $D$ than might be expected.

\subsection{Decoupled modes}

In this sector our results are in remarkable agreement with the numerical ones.
For instance, for the vector mode at $n=100$ and $\ell=2$ we find
\beq
-\text{Im}\,\omega|_{(n=100,\ell=2)}=
\begin{cases} 
1.01044741&\text{numerical \cite{Dias:2014eua}},\\ 
1.01044742&\text{analytical eq.~\eqref{vqnm}},
\end{cases}
\eeq
which is a non-trivial check of the correctness of both calculations.
Fig.~\ref{fig:Vl2} shows that the agreement remains excellent also at smaller values of $n$ and also how the approximation improves with each higher order correction.

\begin{figure}[t]
 \begin{center}
  \includegraphics[width=.5\textwidth,angle=0]{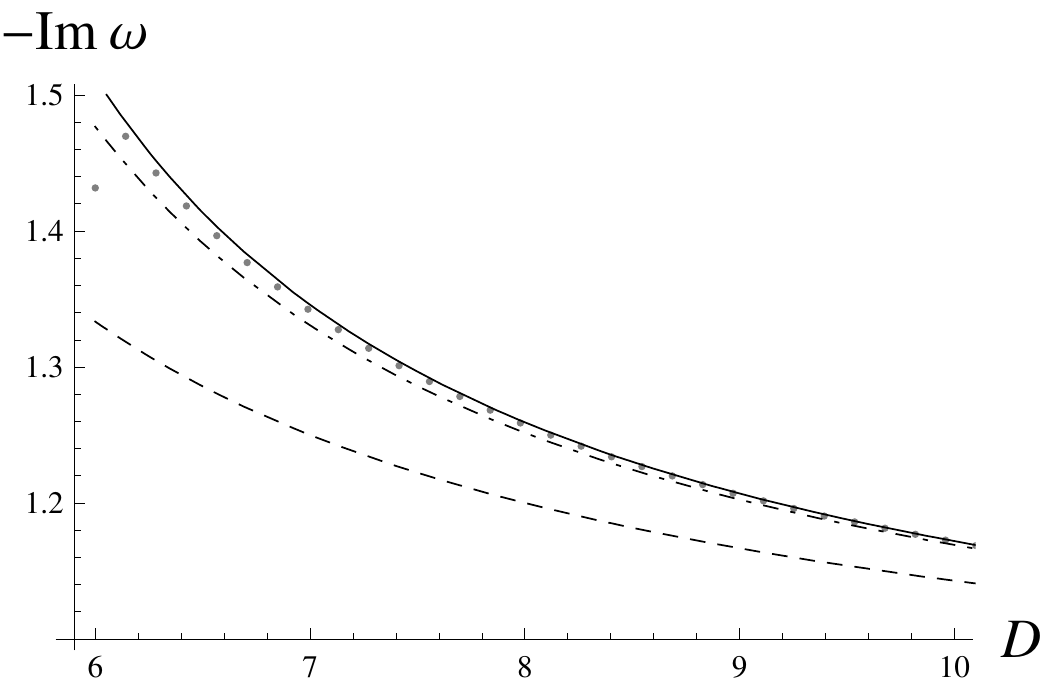}
   \end{center}
 \vspace{-5mm}
 \caption{\small Frequency of decoupled vector quasinormal mode $\ell=2$ as a function of $D$. Solid/dotdashed/dashed lines $=$ eq.~\eqref{vqnm} to $D^{-3}$\,/~to $D^{-2}$\,/~to $D^{-1}$. Gray dots: numerical results \cite{Dias:2014eua}.}
 \label{fig:Vl2}
\end{figure}

At low values of $D$, and in particular at $D=4$ where vector and scalar modes are isospectral, it is not obvious what overtone at a given $\ell$ must be assigned to a decoupled mode obtained in the large $D$ expansion. However, there is one set of modes in  $D=4$ that is  particularly apt for this, namely the algebraically special modes \cite{Chandra:1975}, whose frequency can be computed exactly to be
\beq\label{alsp}
\omega=-i\frac{\ell(\ell^2-1)(\ell+2)}{6}\,.
\eeq
Since these modes are purely imaginary, it is natural to identify them with the decoupled vector modes. The identification does bear out: for the mode $\ell=2$ we find
\beq
-\text{Im}\,\omega|_{(D=4,\ell=2)}=
\begin{cases}
4& \text{exact, eq.~\eqref{alsp}},\\ 
4.25&\mc{O}(1/n^3)\, \text{approximation, eq.~\eqref{vqnm}},
\end{cases}
\eeq
so even at $n=1$, eq.~\eqref{vqnm} approximates the correct value with $6\%$ accuracy. Fig.~\ref{fig:algspec} shows that, although the functional dependence on $\ell$ in \eqref{vqnm} and \eqref{alsp} looks very different, it is nevertheless actually very similar, at least for values of $\ell$ not very much larger than $n$. We find this level of agreement at $n=1$ startling.

\begin{figure}[t]
 \begin{center}
  \includegraphics[width=.5\textwidth,angle=0]{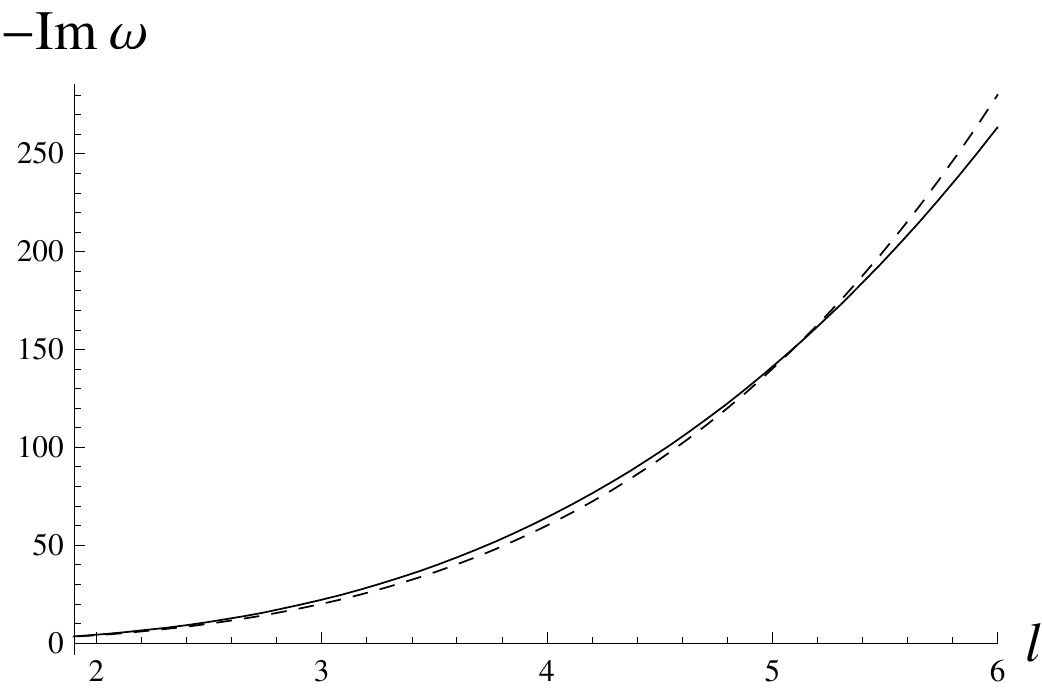}
   \end{center}
 \vspace{-5mm}
 \caption{\small Dashed: frequency \eqref{alsp}, as a function of $\ell$,  of the algebraically special mode of the four-dimensional Schwarzschild black hole. Solid: vector mode frequency \eqref{vqnm} for $D=4$.}
 \label{fig:algspec}
\end{figure}

The accuracy for the scalar modes is also very good, although not as striking as for the vector modes. This could be expected given the more complicated features of the scalar radial potential. In fig.~\ref{fig:Sl2} we present the comparison to \cite{Dias:2014eua} using \eqref{sqnmRe} and \eqref{sqnmIm} for the scalar mode with $\ell=2$. At low $D$ the identification of modes may be complicated or ambiguous due to branch crossings.

\begin{figure}[t]
 \begin{center}
  \includegraphics[width=.47\textwidth,angle=0]{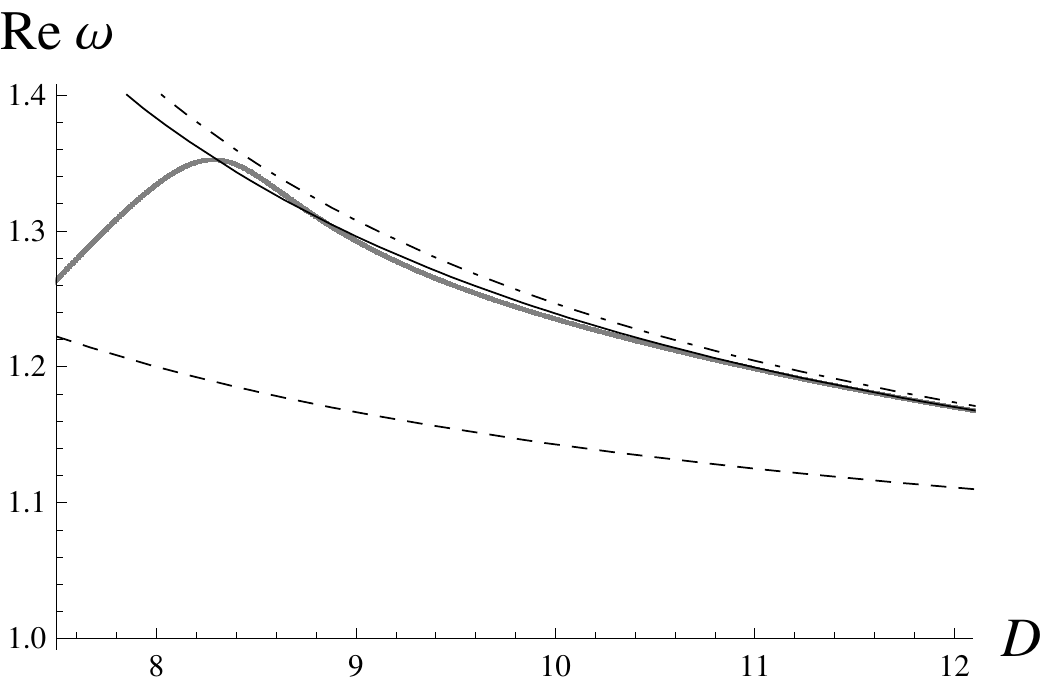}
  \hspace{5mm}
  \includegraphics[width=.47\textwidth,angle=0]{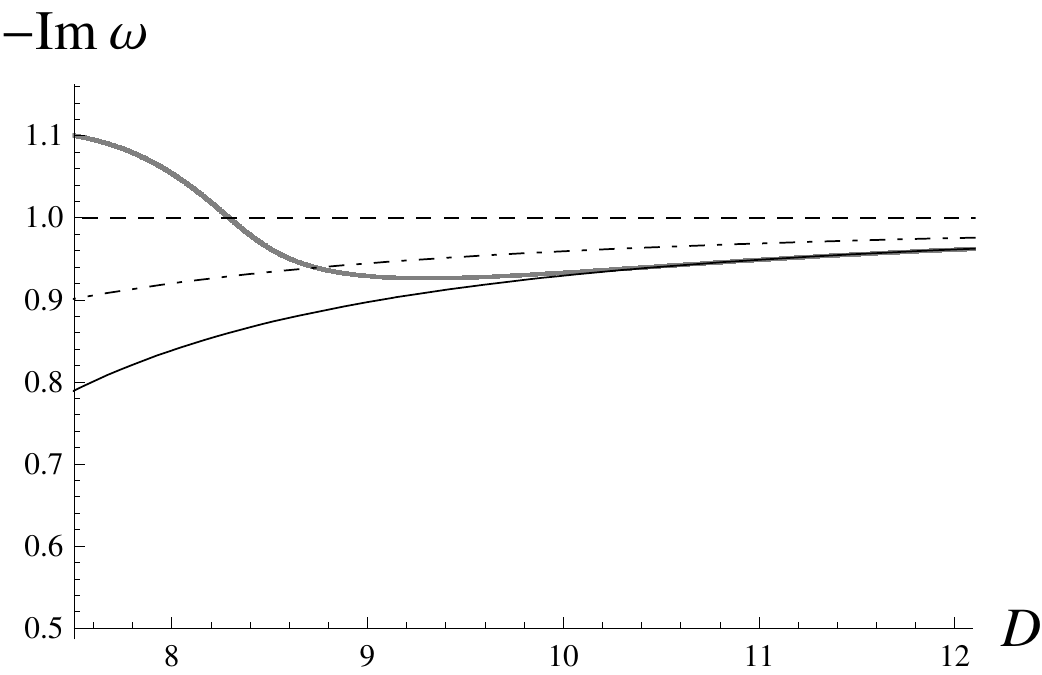}
   \end{center}
 \vspace{-5mm}
 \caption{\small Frequency of (decoupled) scalar quasinormal mode $\ell=2$ as a function of $D$.  Eqs.~\eqref{sqnmRe} (left plot), \eqref{sqnmIm} (right plot) are shown as solid/dotdashed/dashed lines $=$ to $D^{-3}$\,/~to $D^{-2}$\,/~to $D^{-1}$. Gray lines: numerical results \cite{Dias:2014eua}.}
 \label{fig:Sl2}
\end{figure}

\section{Quasinormal modes in the complex frequency plane}
\label{sec:omegaplane}

\begin{figure}[t]
 \begin{center}
  \includegraphics[width=.75\textwidth,angle=0]{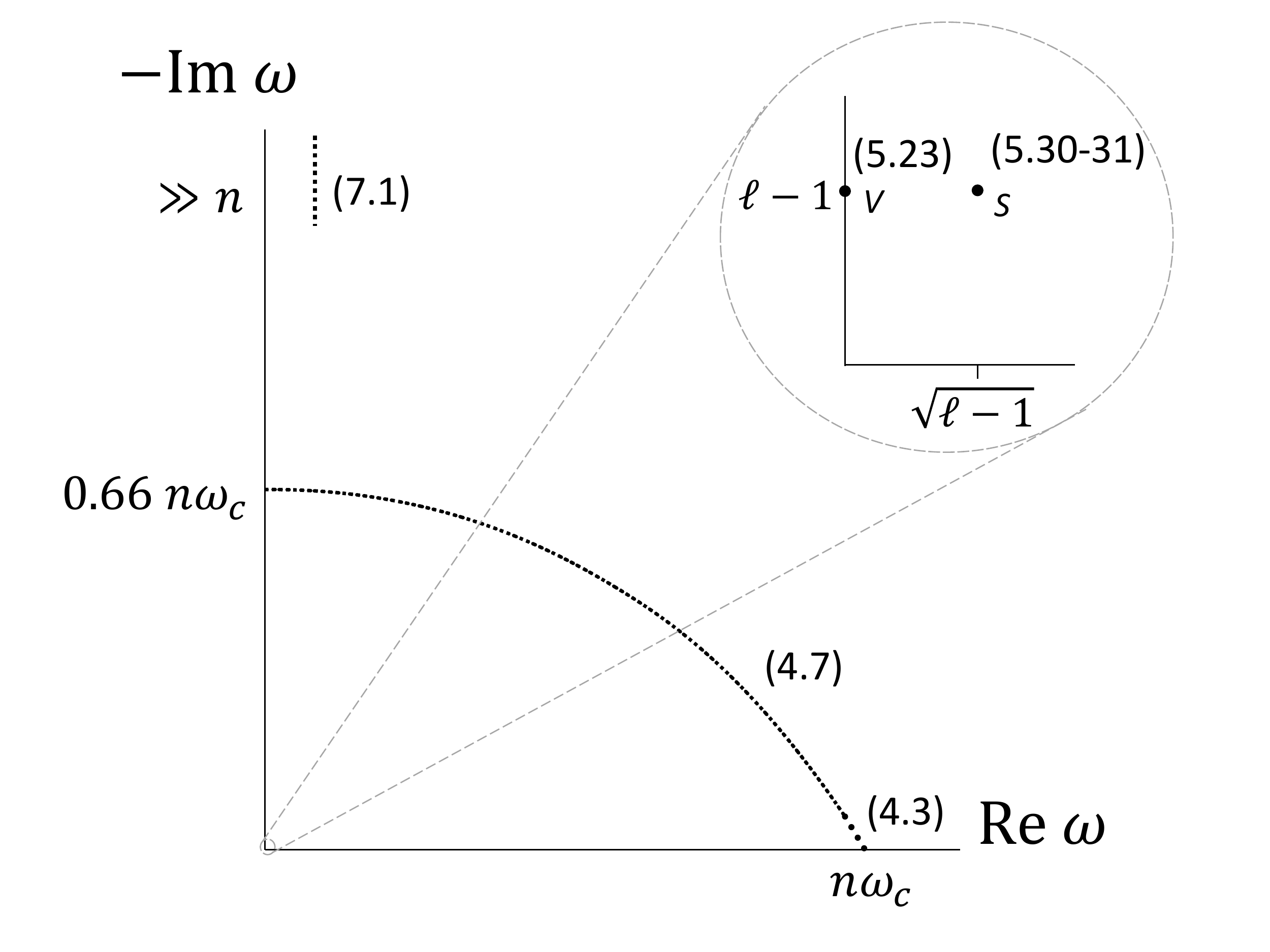}
   \end{center}
 \vspace{-5mm}
 \caption{\small Sketch of large $D$ quasinormal frequencies, for a given value of $\ell$, in the complex $\omega$ plane in units $r_0^{-1}=1$. Dots represent quasinormal frequencies. Scalar, vector, and tensor modes are isospectral except for the vector and scalar decoupled modes close to the origin (present only for $\ell\ll n$), magnified in the inset. 
%The region of $\text{Re}\,\omega \simeq n\omega_c$ and comparatively small $-\text{Im}\,\omega \propto n^{1/3}$ is given by \eqref{onomega}. 
%The dotted curve is the same as in fig.~\ref{fig:ODmodes}, shown here with the expected discretization. 
Highly damped modes are expected in a sequence near the imaginary axis with $-\text{Im}\,\omega \sim n$ and higher, but we have not obtained them in our analysis. Numbers in brackets refer to equations in the text.}
 \label{fig:omegaplane}
\end{figure}

In fig.~\ref{fig:omegaplane} we summarize how the different sets of modes get distributed in the complex frequency plane at large $D$. The main feature is that, for each $\ell$, non-decoupling frequencies are  a distance $\sim D/r_0$ or larger from the origin. Decoupled quasinormal modes, instead, become isolated in a region within a distance of order $1/r_0$ of the origin.

Our large $D$ analysis has focused on the region where $|\omega|^2\leq n^2 \omega_c^2$. But we have not obtained any quasinormal frequency in the range $-\text{Im}\,\omega \gtrsim.6627\, n\omega_c$, since our expansions were not valid for large damping. Such modes must nevertheless exist, and perhaps in some range they can be computed by a refinement of our techniques. Quasinormal modes at extremely large damping have in fact been calculated exactly at finite $n$ \cite{Motl:2002hd,Birmingham:2003rf}, with the result that
\beq\label{monod}
\omega=\frac{n}{2r_0}\lp \frac{\ln 3}{2\pi}-i\lp k+\frac12 \rp\rp\,,
\eeq
where $k$ is the overtone number.
However, since this result is obtained assuming that $k\gg n$, this regime is far from the region of the $\omega$-plane that we have analyzed. At any rate, since these modes are so strongly damped their relevance for the dynamics of black holes is unclear.

\section{Final remarks}
\label{sec:fini}

Our study has centered on the Schwarzschild black hole, but some of its conclusions have broader applicability for general black holes.

The presence of a distinct near-horizon geometry partitions in two the spectrum of black hole oscillations at large $D$. It is tempting to think of the non-decoupling sector as governing the interaction between a black hole and its environment. This interaction appears to be universal for all black holes, in a way perhaps reminiscent of the universality of black hole thermodynamics and the area-law for the entropy. 
Decoupled modes instead contain information specific of each black hole. For instance, the instabilities of black holes at large enough rotations \cite{Emparan:2014jca}, and the hydrodynamic modes of black branes \cite{ESTtba}, appear in this sector.

We expect that these features of the spectrum should be useful for a better understanding of the dynamics of black holes, classical and possibly also quantum.

The analytic determination of the frequency of decoupled modes can be carried very efficiently to high orders in the $1/n$ expansion, with excellent quantitative agreement with numerical computations, in some cases even down to relatively low, realistic dimensions. Although for $D=4$ black holes the method is not competitive in precision with other more developed techniques, its wide applicability may make it convenient for other situations.

\section*{Acknowledgments}

We are very grateful to \'Oscar Dias, Jorge Santos and especially Gavin Hartnett for discussions and for generously supplying the numerical data used in sec.~\ref{sec:numbers}. This work was completed during the workshop ``Holographic vistas on Gravity and Strings'' YITP-T-14-1 at the Yukawa Institute for Theoretical Physics, Kyoto University, whose kind hospitality we acknowledge. While there, we had very useful discussions with Vitor Cardoso, \'Oscar Dias and Paolo Pani.
Work supported by MEC FPA2010-20807-C02-02, AGAUR 2009-SGR-168 and CPAN CSD2007-00042 Consolider-Ingenio 2010. KT was supported by a JSPS grant for research abroad, and by JSPS Grant-in-Aid for Scientific Research No.26-3387.

%\newpage

\addcontentsline{toc}{section}{Appendices}
%\addtocontents{toc}{\protect\setcounter{tocdepth}{0}}
\appendix

\section{Asymptotic expansion of far-zone solution}
\label{app:debye}

When the order and the argument of the Hankel function grow large at the same rate, one can approximate it by the asymptotic formula
\beq\label{hankelasym}
H^{(1)}_\nu(\nu z)=2 e^{-i\pi/3}\lp \frac{4\zeta}{1-z^2}\rp^{1/4} \frac{\text{Ai}\lp e^{i2\pi/3}\nu^{2/3}\zeta\rp}{\nu^{1/3}}\lp 1+\mc{O}\lp(\nu^{-4/3}\rp\rp\,,
\eeq 
where
\beq\label{zetal1}
\frac23\zeta^{2/3}=\ln\lp\frac{1+\sqrt{1-z^2}}{z}\rp-\sqrt{1-z^2}\,,
\eeq
or
\beq\label{zetag1}
\frac23\lp-\zeta\rp^{2/3}=\sqrt{z^2-1}-\text{arcsec}\,z\,.
\eeq
When $z\in\R$, if $z<1$ then \eqref{zetal1} applies, while if $z>1$ \eqref{zetag1} applies instead. The case where $z=1$ gives the expansion in sec.~\ref{sec:homomc}. The function $\zeta(z)$ can be analytically continued in the complex $z$ plane, with a branch cut running along the negative real axis.

The Airy function $\text{Ai}(x)$ in \eqref{hankelasym} can also be expanded for large absolute values of its argument. The expansion we will use is valid when $|\arg\,x|<2\pi/3$, and gives
\beq
\text{Ai}(-x)=\frac1{\sqrt{\pi}\,x^{1/4}}\cos\lp \frac23 x^{3/2}-\frac{\pi}{4}\rp\lp 1-\mc{O}(x^{-3/2})\rp\,.
\eeq

We can now apply these expansions to the far-zone solution \eqref{outmode} in the overlap zone, assuming that $|\hat\omega|^2<\omega_c^2$, to find that
\beq
\Psi_s \propto e^{-n\omega_c r_0 f(\hat\omega/\omega_c)+i\pi/4}\Psi^+(\sR)
+e^{n\omega_c r_0 f(\hat\omega/\omega_c)-i\pi/4}\Psi^-(\sR)\,,
\eeq
where $\Psi^\pm(\sR)$ are the non-normalizable and normalizable wavefunctions in \eqref{psipm}, and $f(z)$ is in \eqref{fz}. This yields \eqref{A+A-far}.

\section{Scalar field solution with $\omega=\mc{O}(n/r_0)$ and $\ell=\mc{O}(n)$}\label{app:scOn}

In the large $n$ limit in the near-zone, the scalar master equation becomes
\begin{eqnarray}
&&0=\Psi_S''(\sR)+\frac{\Psi_S '(\sR)}{\sR-1}\nonum
&&+ \left(\frac{\hat{\omega }^2}{(\sR-1)^2}
-\frac{16 \omega_c^2 \hlam{l}^2 \sR^3-12 \hlam (1+\hlam) \sR^2+\left(8 \hlam+1\right) \sR+1}{4 \sR^2 (\sR-1)  \left(2\hlam \sR+1\right)^2}\right)\Psi_S (\sR)\nonum 
   &&\equiv \cL_{\rm KI}\Psi \label{eq:scalar-hf-heun}
\end{eqnarray}
where we have abbreviated 
\beq
\hlam = \hatl(\hatl+1)=\omega_c^2-\frac14\,,
\eeq
and set $r_0=1$. This is a Heun's differential equation with four singular points at $\sR=0,1,-1/(2\hlam) ,\infty$, which cannot be solved in general. But using a method similar to the one in \cite{Gorbonos:2004uc}, it can be solved through an associated hypergeometric differential equation, namely,
\begin{eqnarray}
 \cL_{\rm Hyp}y(\sR) \equiv y''(\sR) + \left(\fr{\sR}+\frac{1+2i\homega}{\sR-1}\right)y'(\sR)+\frac{i\homega-\hlam}{\sR(\sR-1)}y(\sR).
 \label{eq:scalar-hf-hyp}
\end{eqnarray}
$\cL_{\rm KI}$ and $\cL_{\rm Hyp}$ satisfy the relation
\begin{eqnarray}
\alpha(\sR)^{-1}\cL_{\rm KI}[\alpha(\sR) \cD_1] = \cD_2\cL_{\rm Hyp}
\end{eqnarray}
where
\begin{eqnarray}
 \alpha(\sR)=\frac{(\sR-1)^{-i\homega} \sqrt{\sR}}{1+2\hlam\sR}\,,
\end{eqnarray}
\begin{eqnarray}\label{cD1}
\cD_1 \equiv  \sR(\sR-1) \left(\frac{d}{d\sR}-\frac{\hlam^2+\hat{\omega
   }^2}{2\hlam(1+2\hlam)
   \sR}-\frac{\left(\hlam+i \hat{\omega }\right)^2}{2\hlam(\sR-1)}\right),
%\cD_1 \equiv  \sR(\sR-1) \left(\frac{d}{d\sR}-\frac{\hat{l}^4+2 \hat{l}^3+\hat{l}^2+\hat{\omega}^2}{\left(4 \hat{l}^4+8 \hat{l}^3+6 \hat{l}^2+2 \hat{l}\right)   \sR}-\frac{\left(\hat{l}^2+\hat{l}+i \hat{\omega }\right)^2}{2 \hat{l}   \left(\hat{l}+1\right) (\sR-1)}\right) 
\end{eqnarray}
\begin{eqnarray}
\cD_2\equiv \sR(\sR-1)\left(\frac{d}{d\sR}+\frac{\hlam(7\hlam+4)-\hat{\omega }^2}{2\hlam(1+2\hlam) \sR}
-\frac{\hlam(\hlam-4)+2i\hlam\homega-\homega^2}{2 \hlam (\sR-1)}
-\frac{4 \hlam}{2\hlam\sR+1}\right).
 \end{eqnarray}
If $y(\sR)$ is a solution of eq.~(\ref{eq:scalar-hf-hyp}), then $\Psi(\sR) = \alpha(\sR) \cD_1 y(\sR)$ becomes a solution of eq.~(\ref{eq:scalar-hf-heun}).
The ingoing solution can now be found to be given by \eqref{nhs0}.

\section{Universal spectrum in an extremal charged black hole}
\label{app:extr}

For illustration we consider a massless scalar field propagating in the geometry of the $D$-dimensional extremal Reissner-Nordstrom solution. The same field equation describes its gravitational tensor perturbations. 

The far-zone solution is the same as in sec.~\ref{sec:farnear}. The limiting near-horizon geometry at large $D$ can be obtained from the general analysis in \cite{Emparan:2013xia}, with the two-dimensional $(\hat t, \sR)$ sector being
\beq
n^2 ds^2_\text{nh}=-\lp 1-\frac1{\sR}\rp^2 d\hat t^2+\frac{d\sR^2}{(\sR-1)^2} 
\eeq
(we set the horizon radius $r_0=1$). The field equation is
\beq
\frac{d}{d\sR}\lp (\sR-1)^2\frac{d}{d\sR}\Psi\rp +\frac{\sR^2 }{(\sR-1)^2}\hat\omega^2\Psi-\lp\omega_c^2-\frac14\rp\Psi=0\,.
\eeq
The ingoing condition at the future horizon requires that
\beq
\Psi \sim e^{-i\hat\omega/(\sR-1)}
\eeq
near $\sR=1$, and the solution that satisfies it is given in terms of a Whittaker $W$ function
\beq
\Psi=W\lp i\hat\omega, \sqrt{\omega_c^2-\hat\omega^2},\frac{2i\hat\omega}{\sR-1}\rp\,.
\eeq
At large $\sR$ this solution contains normalizable and non-normalizable components $\Psi^{\pm}$ with ratio $|A_+/A_-|=\mc{O}(1)$. When $\hat\omega=\omega_c$ one obtains \eqref{ABform} with $|A/B|=\mc{O}(1)$. Since these are the same conditions as are used in sec.~\ref{sec:nondec}, we obtain the same, universal, non-decoupling spectrum.

\section{Scalar quasinormal frequencies with $\omega=\mc{O}(1/r_0)$ and $\ell=\mc{O}(1)$}\label{app:scqnm}

\subsection{Master variable formulation: next to leading order}\label{app:scmaster}

If we include the next-to-leading order, the solution for $\Psi_S(\sR)$ with the horizon ingoing boundary condition is
\begin{eqnarray}
\Psi_S(\sR) &=& \Psi_S^{(0)}(\sR)+\fr{n} \Psi_S^{(1)}(\sR)\nonum\\
&=&
\sqrt{\sR}\left[1+\frac{1}{n}\left( ( 1-2 \ell+2 i \omega) \ln \sqrt{\sR}-2 (\ell-1) (\sR-1)-i \omega  \ln (\sR-1)\right)\right],
\end{eqnarray}
where we have fixed the overall amplitude like in \eqref{hborders}.
The expansion of $\Psi_S(\sR)$ at large $\sR$ gives
\begin{eqnarray}\label{lpsi}
\Psi_S(\sR) &=&  \sqrt{\sR}\left[1 + \frac{1}{n}\left(\frac{i\omega}{\sR}+2(\ell-1)-2(\ell-1) \sR -(2\ell-1) \ln\sqrt{\sR} +\ord{\sR^{-2}}\right)\right]\nonum&&+\ord{n^{-2}}\,.
\end{eqnarray}
Here we expanded the terms at order $n^{-1}$ only up to $\sR^{-1}$, since  $\sR \ll n$ in the matching region.

To the same order, the solution for $\bar\Psi_S(\bar{\sR})$ with the condition $\bar\Psi_S(\bar{\sR}) \sim \bar{\sR}^{-1/2}$ as $\bar{\sR}\rightarrow \infty$ is
\begin{eqnarray}
\bar\Psi_S^{(0)}(\bar{\sR})+\frac{1}{n}\bar\Psi_S^{(1)}(\bar{\sR})&=& \frac{B_0\sqrt{\bar{\sR}}}{1+2(\ell-1)\bar{\sR}} \left[1-\frac1{n}\left(\frac{3+(2-6\ell+4\ell^2)\bar{\sR}}{2+4(\ell-1)\bar{\sR}}-(2\ell-1)\ln \sqrt{\bar{\sR}} \right)\right].\nonumber\\
\end{eqnarray}
Matching the leading order amplitude requires $B_0 = \sqrt{n}  + \ord{n^{-1/2}}$. If we write $B_0 = \sqrt{n}\lp 1+B_1/n \rp$,
the expansion in $1/n$ becomes
\begin{eqnarray}
\bar\Psi_S^{(0)}(\bar{\sR})+\frac{1}{n}\bar\Psi_S^{(1)}(\bar{\sR})= \sqrt{\sR}\left[1+\frac{1}{n}\left(b_1+b_2 \sR+b_3 \ln \sR\right)+\ord{n^{-2}}\right],
\end{eqnarray}
where
\begin{eqnarray}
b_1=B_1+ \lp\ell- \frac{1}{2}\rp \ln n-\frac32 ,\qquad b_2=-2 (\ell-1) ,\qquad b_3=\frac{1}{2}-\ell\,.
\end{eqnarray}
However, this is not enough to do the matching, since a term $\sim1/\bar{\sR}$ in $\bar\Psi_S^{(2)}(\bar\sR)$ would also contribute to order $1/n$. We find that there is such a term,
\begin{eqnarray}
  \frac{1}{n^2}\bar\Psi_S^{(2)}(\sR/n) = \frac{1}{n}\frac{\ell-\ell^2+\omega^2}{2(\ell-1)\sR}+\ord{n^{-2}}
\end{eqnarray}
(we do not show the other terms in $\bar\Psi_S^{(2)}(\bar{\sR})$ that we do not need).
Then, the correct expansion up to next-to-leading order is 
\begin{eqnarray}\label{tilpsi}
\bar\Psi_S(\sR)\simeq \sqrt{\sR}\left[1+\frac{1}{n}\left(\frac{b_0}{\sR}+b_1+b_2 \sR+b_3 \ln \sR\right)+\ord{n^{-2}}\right],
\end{eqnarray}
where
\begin{eqnarray}
&&b_0=\frac{ \ell^2-\ell-\omega ^2}{2(\ell-1)},
\end{eqnarray}
Matching \eqref{tilpsi} and \eqref{lpsi} requires
\begin{eqnarray}
 B_1&=& 2\ell-\fr2- \lp\ell- \fr2\rp \ln n,\nonum 
i\omega&=& \frac{ \ell^2-\ell-\omega ^2}{2(\ell-1)}\,.
\end{eqnarray}
This last equation gives the quasinormal frequencies \eqref{omsc0}.

\subsection{Higher order calculations}\label{app:schigher}
Using the Kodama-Ishibashi gauge-invariant variables $X,Y,Z$~\cite{Kodama:2003jz}, the leading equation decouples for
\begin{eqnarray}
 X(\sR)=\fr{2}P(\sR)+\fr{2}\frac{\sR}{\sR-1} Q(\sR),\quad  Y(\sR)=\fr{2}P(\sR)-\fr{2}\frac{\sR}{\sR-1} Q(\sR)
\end{eqnarray}

\begin{eqnarray}
  \frac{d}{d\sR} P^{(k)}(\sR) = \cS^{(k)}_{P},\quad   \frac{d}{d\sR} Q^{(k)}(\sR) = \cS^{(k)}_{Q}, \label{eq:ginv-dev}
\end{eqnarray}
and
\begin{eqnarray}
Z^{(k)}(\sR) = -\fr{2\ell}(P^{(k)}(\sR)+Q^{(k)}(\sR)) + \cS^{(k)}_{Z},\label{eq:ginv-z}
\end{eqnarray}
where variables are expanded in $1/n$ as
\begin{eqnarray}
  X(\sR) = \sum_{k\geq 0}\frac{X^{(k)}(\sR)}{n^k},\quad  Y(\sR) = \sum_{k\geq 0}\frac{Y^{(k)}(\sR)}{n^k},\quad  Z(\sR) = \sum_{k\geq 0}\frac{Z^{(k)}(\sR)}{n^k},
\end{eqnarray}
and
\begin{eqnarray}
  P(\sR) = \sum_{k\geq 0}\frac{P^{(k)}(\sR)}{n^k},\quad  Q(\sR) = \sum_{k\geq 0}\frac{Q^{(k)}(\sR)}{n^k}.
\end{eqnarray}
The differential equation for $Z^{(k)}(\sR)$ is automatically satisfied if eqs.~(\ref{eq:ginv-dev}) and (\ref{eq:ginv-z}) hold, by virtue of the Bianchi identity.
The leading solution becomes
\begin{eqnarray}
 P^{(0)} = \ell P_0,\quad Q^{(0)} = \ell Q_0.
\end{eqnarray}
Since the integration constants that appear at higher order can be absorbed by a redefinition of $P_0,Q_0$, the amplitude of the perturbation is determined only by $P_0$ and $Q_0$.

In the original variables this is
\begin{eqnarray}
  X^{(0)} = \frac{\ell P_0}{2}+\frac{\ell Q_0 \sR}{2(\sR-1)},\quad   Y^{(0)} = \frac{\ell P_0}{2}-\frac{\ell Q_0 \sR}{2(\sR-1)},\quad
   Z^{(0)} = -\fr{2}(P_0+Q_0).
\end{eqnarray}

At the next order, the solutions for $P,Q$ are
\begin{eqnarray}
 P^{(1)}(\sR)&=&\sR \left(\ell^2 \left(P_0-Q_0\right)+\ell  \left(Q_0-P_0\right)-\omega ^2
   \left(P_0+Q_0\right)\right)\nonum
   &&\qquad-\ln (\sR-1) \left(\ell ^2 Q_0+\omega ^2
   \left(P_0+Q_0\right)\right)-\ell  P_0 \ln \sR,\\
Q^{(1)}(\sR)&=&\sR \left(\ell ^2 \left(P_0-Q_0\right)+\ell  \left(Q_0-P_0\right)-\omega ^2
   \left(P_0+Q_0\right)\right)\nonum
   &&\qquad -\ell  \ln \sR \left(\ell  P_0+Q_0\right)-\frac{\ell 
   P_0}{\sR}+\omega ^2 \left(P_0+Q_0\right) \ln (\sR-1).
\end{eqnarray}
We have also computed the solutions at second, third and fourth order, but they are too long to give explicitly here.

\subsubsection{Boundary condition}
The ingoing condition is imposed at $\sR=1$, in such a way that the following quantities are regular,
\begin{eqnarray}
 &&\left( X^{(0)}+\fr{n}X^{(1)}+\fr{n^2}X^{(2)}+\fr{n^3}X^{(3)}+\fr{n^4}X^{(4)}\right)\times (\sR-1)^{1+i\omega r_0/n}\nonum
 &=&\Bigl(\alpha_4(\omega)+ a_4(\omega) \ln (\sR-1)+b_4(\omega) \ln^2(\sR-1)+c_4(\omega)\ln^3(\sR-1)+d_4(\omega)\ln^4(\sR-1)
\nonum
&&+\ord{n^{-5}}\Bigr)+\ord{\sR-1}
\end{eqnarray}
where it turns out that $a_4(\omega)=0$ is the only independent condition and
\begin{eqnarray}
a_4(\omega)&=&\frac{\omega}{2n}\Bigl[ P_0 \omega+Q_0  (\omega +i\ell )  \nonum
&&\quad+\ \frac{i P_0  \left(\ell ^2+\ell  (-2-i \omega )-\omega  (\omega -3   i)\right)- i Q_0  \left(\ell ^2+\ell  (-1-i \omega )+\omega   (\omega +3 i)\right)}{n}\nonum
   &&+\ \frac{1}{\ell n^2}\Bigl(P_0  \left(-i \ell ^4-\ell ^3 (\omega -2 i)+\ell ^2 (4-i \omega ) \omega +\ell  \omega  \left(2 \omega ^2+2 i \omega
   -3\right)-\omega ^3\right)\nonum
   &&\qquad+\ Q_0  \left(i \ell ^4+\ell ^3 (3 \omega
   -2 i)+i \ell ^2 \left(3 \omega ^2+6 i \omega +1\right)+\ell  \omega  \left(2 \omega
   ^2-4 i \omega +3\right)-\omega ^3\right)\Bigr)\nonum
   &&+\ \frac{1}{\ell n^3}\Bigl(i P_0
    \left(\ell ^5+\ell ^4 (-3-3 i \omega )+\ell ^3 \left(\omega ^2+11 i \omega +3\right)-\ell ^2
   \left(6 \omega ^2+11 i \omega +2\right)\right.\nonum
   &&\left.\hspace{6cm} +\ \ell  \omega  \left(\omega ^3+4 i \omega^2+5 \omega +3 i\right)+\omega ^4\right)\nonum
   &&\qquad +\ Q_0 \left(-i  \ell ^5+\ell ^4 (-5 \omega +3 i)+\ell ^3 \left(-5 i \omega ^2+11 \omega -3 i\right)\right.\nonum
   &&\qquad\qquad\quad+\ell ^2
   \left(-8 \omega ^3+12 i \omega ^2-9 \omega +i\right)\nonum
   &&\left.\qquad\qquad\quad+\ell  \omega  \left(i \omega
   ^3+12 \omega ^2-7 i \omega +3\right)+i \omega ^3 (\omega +6 i)\right)\Bigr)\Bigr].
   \end{eqnarray}
The asymptotic boundary condition is imposed by requiring
\begin{eqnarray}
 && \left( X^{(0)}+\fr{n}X^{(1)}+\fr{n^2}X^{(2)}+\fr{n^3}X^{(3)}+\fr{n^4}X^{(4)}\right)/\sR\nonum
  &&\hspace{4cm} \simeq \tilde{a}_4(\omega)+\tilde{b}_4(\omega)\ln \sR+\tilde{c}_4(\omega)\ln^2 \sR+\tilde{d}_4(\omega)\ln^3 \sR+\ord{n^{-5}} =0. \nonum
\end{eqnarray}
Similarly, $\tilde{a}_4(\omega)=0$ is the only independent condition,
\begin{eqnarray}
&&\tilde{a}_4(\omega)=\frac{P_0 \left(\ell ^2-\ell -\omega ^2\right)+Q_0 \left(-\ell ^2+\ell -\omega
   ^2\right)}{n}\nonum
   &&\quad+\ \frac{P_0 \left(-\ell ^3+2 \ell ^2-\ell  \left(\omega ^2+1\right)+2 \omega
   ^2\right)+Q_0 \left(\ell ^3-2 \ell ^2+3 \ell  \omega ^2+\ell -4 \omega
   ^2\right)}{n^2}\nonum
   &&\quad+\ \frac{1}{6\ell n^3}\Bigl(P_0 \left(\left(6+\pi ^2\right) \ell ^5-\left(18+\pi^2\right) \ell ^4+2 \ell ^3 \left(\left(6+\pi ^2\right) \omega ^2+9\right)\right.\nonum
    &&\hspace{5cm}\left.-\ell ^2  \left(7 \left(6+\pi ^2\right) \omega ^2+6\right)+\ell  \omega ^2 \left(\pi ^2   \left(\omega ^2+4\right)+30\right)+6 \omega ^4\right)\nonum
   &&\qquad+\ Q_0   \left(\left(\pi ^2-6\right) \ell ^5-\left(\pi ^2-18\right) \ell ^4+2 \ell ^3  \left(\left(\pi ^2-18\right) \omega ^2-9\right)\right.\nonum
   &&\hspace{5cm}\left.+\ell ^2 \left(\left(78-5 \pi^2\right) \omega ^2+6\right)+\ell  \omega ^2 \left(\pi ^2 \left(\omega
   ^2+4\right)-42\right)+6 \omega^4\right)\Bigr)\nonum
   &&+\frac{1}{6\ell n^4}\Bigl[P_0  \Bigl(\ell ^6 \left(-12 \zeta (3)-6+\pi ^2\right)-4 \ell ^5 \left(\pi ^2-3 (\zeta
   (3)+2)\right)\nonum
   &&\qquad\quad-\ \ell ^4 \left(\pi ^2 \left(2 \omega ^2-3\right)+12 \left(2 \omega ^2 (\zeta (3)+1)+3\right)\right) +2 \ell ^3 \left(\omega ^2 \left(42 \zeta(3)+63+\pi ^2\right)+12\right)\nonum
   &&\qquad\quad\left.-\ \ell ^2 \left(3 \omega ^4 \left(4 \zeta (3)-4+\pi^2\right)+\omega ^2 \left(120 \zeta (3)+150-7 \pi ^2\right)+6\right)\right.\nonum
   &&\qquad\quad +\ 2 \ell  \omega ^2 \left(2 \pi ^2 \left(\omega ^2-2\right)+3 \left(\omega ^2 (4 \zeta
   (3)-5)+8 (\zeta (3)+1)\right)\right)+6 \omega ^4\Bigr)\nonum
   &&\qquad+\ Q_0  \Bigl(-\left(\pi ^2-6\right) \ell ^6+2 \left(\pi ^2-12\right) \ell ^5-\ell ^4 \left(6 \pi
   ^2 \omega ^2-72 \omega ^2+\pi ^2-36\right)\nonum
   &&\qquad\quad+\ 2 \ell ^3 \left(\omega ^2 \left(12  \zeta (3)-99+11 \pi ^2\right)-12\right) \nonum
   &&\qquad\quad+\ell ^2 \left(\left(12-5 \pi ^2\right) \omega ^4+\omega ^2 \left(-72 \zeta (3)+186-31 \pi ^2\right)+6\right)  \nonum
  &&\qquad\quad+\ 2 \ell 
   \omega ^2 \left(\pi ^2 \left(5 \omega ^2+8\right)+3 \left(\omega ^2 (4 \zeta (3)-9)+8 \zeta (3)-10\right)\right)+42 \omega ^4\Bigr)\Bigr].
\end{eqnarray}
The same conditions are obtained from the regularity of $Y$ and $Z$.

The equations $a_4=0$ and $\tilde{a_4}=0$ admit a nontrivial solution for $P_0$ and $Q_0$ if and only if
\begin{eqnarray}
&& (2-2 \ell ) \omega -i (\ell -1) \ell +i \omega ^2+\frac{i (\ell -1) \ell ^2+i (\ell -1) \omega ^2+2 (\ell -1)^2
   \omega +2 \omega ^3}{n}\nonum
   &&+\  \frac{1}{n^2}\Bigl(-\frac{1}{6} i \left(2 \left(6+\pi ^2\right) \ell ^2-\left(24+7 \pi ^2\right)
   \ell +4 \left(3+\pi ^2\right)\right) \omega ^2\nonum
   &&\qquad\quad-\frac{1}{6} i (\ell -1) \ell  \left(\left(6+\pi ^2\right) \ell ^2-6 \ell +6\right)-\ \frac{i \left(\pi ^2 \ell +6\right)
   \omega ^4}{6 \ell }\nonum
   &&\qquad\quad+\left(-6 \ell -\frac{2}{\ell }+\frac{\pi ^2}{3}+8\right) \omega ^3-2
   (\ell -1)^2 (2 \ell -1) \omega \Bigr)\nonum
   &&+\ \frac{1}{n^3}\Bigl[\frac{i \omega ^4 \left(\ell ^2 \left(12 (\zeta
   (3)-1)+5 \pi ^2\right)-3 \ell  \left(8 \zeta (3)-10+3 \pi ^2\right)-12\right)}{6
   \ell }\nonum
   &&\qquad+\ \omega^3 \left(4 \ell ^2 (\zeta (3)+4)-\frac{2}{3} \ell  \left(15 \zeta(3)+54+\pi ^2\right)-\frac{8}{\ell }+8 \zeta (3)+\pi ^2+28\right)\nonum
      &&\qquad+\ \frac{1}{3} i \omega ^2 \left(3 \ell ^3 \left(4 \zeta (3)+4+\pi^2\right)-\ell ^2 \left(42 \zeta(3)+45+10 \pi ^2\right)\right.\nonum
      &&\hspace{5cm}+\ \left.\ell  \left(60 \zeta (3)+45+9 \pi ^2\right)-2 \left(12 \zeta (3)+6+\pi  ^2\right)\right)\nonum
   &&\qquad+\ \frac{1}{3} (\ell -1) \omega  \left(6 \ell ^3 (\zeta (3)+3)-\left(30+\pi ^2\right) \ell ^2+18 \ell -6\right)+\omega ^5 \left(2 \zeta (3)-\frac{2}{\ell }\right)\nonum
   &&\qquad+\ \frac{1}{6} i  (\ell -1) \ell ^2 \left(\ell ^2 \left(12 \zeta (3)+6+\pi ^2\right)-12  \ell +12\right)
\Bigr]=0.
\end{eqnarray}
From here we obtain \eqref{sqnmRe} and \eqref{sqnmIm}.

%%%%%%%%%%%%%%%%%%%%%%%%%%%%%%%%%%%%%%%%%

\end{document}